\documentclass{emulateapj}
\usepackage{graphicx}
\usepackage{epstopdf}
\def\msun{\ifmmode {\rm M}_{\mathord\odot}\else $M_{\mathord\odot}$\fi}
\newcommand{\nbh}	{\bar n_{\rm H}}
\newcommand{\nbht}	{\bar n_{\rm H,\, 3}}
\newcommand{\e}	{\ifmmode ^{-1}\else $^{-1}$\fi}
\newcommand{\ee}	{\ifmmode ^{-1}\else $^{-2}$\fi}
\newcommand{\eee}	{\ifmmode ^{-1}\else $^{-3}$\fi}

\begin{document}

\title{Driven and Decaying Turbulence Simulations of Low-Mass Star Formation: From Clumps to Cores to Protostars}

\author{Stella S. R. Offner} 
\affil{Department of Physics, University of
California, Berkeley} 
\email{soffner@berkeley.edu}
\author{ Richard I. Klein}
\affil{Department of Astronomy, University of California Berkeley, Berkeley 
CA 94720, USA, and Lawrence Livermore National Laboratory, P.0. Box 808, L-23,
Livermore, CA 94550, USA}
\author{Christopher F. McKee}
\affil{Departments of Physics and Astronomy, University of California Berkeley, Berkeley, CA 94720, USA}

\begin{abstract}
Molecular clouds are observed to be turbulent, but the origin of this turbulence is not
well understood. As a result, there are two different approaches to simulating
molecular clouds, one in which the turbulence is allowed to decay after it is initialized,
and one in which it is driven.
We use the adaptive mesh refinement (AMR) code, Orion, to perform high-resolution simulations of molecular cloud cores and protostars in environments 
with both driven and decaying turbulence.
We include self-gravity, use a barotropic equation of state, and represent regions exceeding the maximum grid resolution with sink particles. We analyze the properties of bound cores such as size, shape, linewidth, and rotational energy, and we find reasonable agreement with observation.  At high resolution, the different rates of core accretion in the two cases have a significant effect on protostellar system development. Clumps forming in a decaying turbulence environment produce high-multiplicity protostellar systems with Toomre-Q unstable disks that exhibit characteristics of the competitive accretion model for star formation. In contrast, cores forming in the context of continuously driven turbulence and virial equilibrium form smaller protostellar systems with fewer low-mass members. 
Our simulations of driven and decaying turbulence
show some statistically significant differences, particularly in the
production of brown dwarfs and core rotation, but the uncertainties are large enough
that we are not able to conclude whether observations favor one or the
other.

\end{abstract}

\section{ INTRODUCTION}

Contemporary star formation occurs exclusively in dense molecular clouds (MCs). Such regions exhibit 
large non-thermal line widths generally attributed to supersonic turbulence (Larson 1981). Although 
debate continues on the origin and characteristics of this turbulence, it is now 
recognized that turbulence is a necessary element of star formation and plays an important role in the shape of the core initial mass function (IMF), the lifetimes of molecular clouds, and the star formation rate.

Simulations have shown that supersonic turbulence decays with an e-folding time of approximately one cloud crossing time if there is no energy input to sustain it (Stone et al. 1998; Elmegreen \& Scalo 2004; Mac Low \& Klessen  2004). 
If turbulence decays as quickly in molecular clouds then star formation must happen rapidly as the cloud looses turbulent pressure support and undergoes global collapse. In this senario, star formation occurs on a dynamical timescale and MCs must be transient dynamic structures (Elmegreen 2000; Hartmann 2001). If however, turbulence is fed from large scales or protostellar winds, 
expanding HII regions, and other processes provide sufficient energy injection to balance dissipation produced by shocks,
then MCs may arrive at a quasi-equilibrium state 
(e.g., Tan, Krumholz \& McKee 2006).
Although there are many possible sources for turbulent energy, the dominant source and the specific characteristics of turbulence remain poorly understood. Recent effort has been directed at this issue and some observations of low-mass star forming regions, e.g. L1551, find evidence for ongoing turbulence injection in the form of winds and jets, which maintains rough virial balance over several cloud dynamical times (Swift \& Welch 2007). While turbulent support is maintained, only a small number of overdense regions will become gravitationally unstable and form stars in a dynamical time, leading to a low star formation rate and allowing MCs to live for several dynamical times.

The two different views of cloud dynamics are related to, but distinct from, the two
major approaches to simulating turbulent molecular clouds. 
The fact that
there are two competing approaches to the simulation of such clouds is a direct 
reflection of our lack of understanding of the origin of the turbulence in these clouds
(McKee \& Ostriker 2007). 
In one method, the turbulence is initialized
and then allowed to decay (e.g. Klessen et al. 1998; Bonnell et al. 2003;  Bate et al. 2002; Tilley \& Pudritz 2004). 
The primary problem with this approach
is that the turbulence decays to levels that are much lower than those observed. Advocates
of this method argue that the gravitational collapse that ensues after the decay of
the turbulence can be observationally confused with turbulence
(e.g. Vazquez-Semadeni et al. 2006), but it is difficult to see
how to maintain a low star formation efficiency if much of the gas is in a state of
gravitational collapse. The other approach to cloud simulation is to drive the turbulence so
that it does not decay (e.g. Padoan \& Nordlund 1999; Gammie et al. 2003; Li et al. 2004; Jappsen et al. 2005).
This approach allows one to
study the processes that occur at a given level of turbulence, which can be set to
match that in a given cloud, but it suffers from the disadvantage that the driving is
unphysical. This approach is a good match for the case of quasi-equilibrium clouds,
and it can be made consistent with the case of transient clouds if it is assumed that the
simulation box represents only a small part of the molecular cloud, so that the decay  time
for the turbulence is long compared to the dynamical time of the simulation. 

The near universal shape of the stellar IMF across diverse star forming 
environments has sparked much debate and generated diverse theories.   
Padoan \& Nordlund (2004) suggest that the functional form of the IMF can be derived from 
the power spectrum and probability density function characteristic of supersonic turbulence.
Larson (2005) proposes that the peak of the IMF is set by the Bonnor-Ebert mass at the minimum cloud temperature, which is related to the dust-gas coupling and gas cooling efficiency.
In the competitive accretion model, Bonnell et al. (2004) invoke high stellar densities at the centers of clusters to propose that the relative position of the stars in the gas reservoir determines their mass. According to this model, the IMF is determined by mass segregation, such that low-mass stars form in the lower density gas at the edges of the cluster and higher mass stars form in the centers, where their masses can be further increased by the coalescence of smaller protostars.
In addressing the origin of the IMF, numerical simulations have been largely inconclusive in discriminating between models given that a wide range of conditions (virial parameters, resolution, code algorithms, included physics) have all succeeded in reproducing the IMF shape.

A large amount of computational effort has been directed towards studying 
self-gravitating turbulent clouds both with and without magnetic fields (e.g.  Klessen 2001; Bonnell et al. 2003; Li et al. 2004).
A number of simulations succeed in reproducing various observed core properties such as the 
IMF and Larson's laws (Padoan \& Nordlund 1999; Gammie et al. 2003; Tilley \& Pudritz 2004; Li et al. 2004; Jappsen et al. 2005; Bate \& Bonnell 2005).  
However, most simulations lack the resolution to span the turbulent inertial range (Klein et al. 2007)
to accurately render the evolution of cores into stars in a cluster environment. 

In this paper, we perform numerical AMR simulations with our code Orion to investigate the role of driven and decaying turbulence on low-mass star formation. 
We follow the evolution of star forming cores in a turbulent box to show that turbulent feedback is correlated with
the multiplicity of stellar systems, the shape of the IMF,  and the dominant protostellar accretion model. In $\cal x$2, we discuss the methodology of Orion and the initial conditions. In $\cal x$3, we analyze core properties in driven and decaying turbulence at low resolution. In $\cal x$4, we present results from a few high resolution studies of the protostellar core evolution inside selected cores formed in the context of driven and decaying turbulence. Finally, $\cal x$5 contains conclusions.  
In a companion paper (Offner et al. 2008, hereafter Paper II) we investigate the effects of
driven and undriven turbulence on the properties of the cores from which the stars form.

\section { CALCULATIONS}
\subsection{ { Numerical Methods}}

Our simulations are performed using the parallel adaptive mesh refinement (AMR) code, Orion,
which uses a conservative second order Godunov scheme to solve the equations of 
compressible gas dynamics (Truelove et al. 1998; Klein 1999). 
Orion solves the Poisson equation using 
multi-level elliptic solvers with multi-grid iteration. 
Throughout our calculations, we use the Truelove criterion to determine the addition of new AMR grids (Truelove et al. 1997),
\begin{equation}
{\rho} < {\rho_{\rm J}} = {{J^2\pi c_{\rm s}^2}\over{G(\Delta x^l)^2}},
\end{equation}
where $\Delta x^l$ is the cell size on level, $l$, and we adopt a Jeans number of $J=0.25$.
We insert sink particles 
in regions of the flow that have exceeded this density on the maximum level 
(Krumholz et al. 2004). Sink particles serve as numerical markers of collapsing regions and also, after sufficient mass accretion and lifetime, protostellar objects.  We impose a merger criterion that combines sink particles that approach within two grid cells of one another but prohibits nearby objects from merging if both have masses exceeding 0.1 $\msun$. This limit divides stars from substellar objects such as brown dwarfs and has the effect of tracking all significantly massive objects. Particles that represent temporary violations of the Jeans condition and have little bound mass tend to accrete little and ultimately merge with their more substantial neighbors.
The combination of sink particles and AMR with the Jeans criterion allows us to accurately and efficiently continue our calculation to high resolution without the time constraints imposed by 
a large base grid size and without the consequences of artificial fragmentation.

\subsection { Initial Conditions and Simulation Parameters}

Isothermal self-gravitating gas is scale free, so we give the key cloud properties as a function
 of fiducial values for the mean number density of hydrogen nuclei, 
$\nbh$, and gas temperature, $T$. We chose a characteristic
3-D turbulent Mach number, ${{\cal M}}$=8.4, such that the cloud
is approximately virialized:
\begin{equation}
{ \alpha_{\rm vir}} = {{5 \sigma^2} \over { G M / R}} \sim 1.67. 
 \end{equation}
It is then easy to scale the simulation results to the 
astrophysical region of interest. 
For the adopted values of the virial parameter and Mach number,
the box length, mass, and 1-D velocity dispersion are given by
\begin{eqnarray}
L &=&  \mbox{2.9 } {T_1 }^{1/2} {\bar n_{\rm H, 3}}^{-1/2}  \mbox{ pc}\, ,
\label{eq:l} \\
M &=&  \mbox{865 } {T_1}^{3/2} {\bar n_{\rm H, 3}}^{-1/2} \mbox{ M}_{\odot} \, ,
\label{eq:m}\\
\sigma_{\rm 1D} &=&  \mbox{0.9 } {T_{1}}^{1/2} \mbox{ km s}^{-1}\, ,\\
t_{\rm ff} &=&  \mbox{1.37 } {\bar n_{\rm H,3}}^{-1/2} \mbox{ Myr}\, ,
\end{eqnarray}
where $\nbht=\nbh/(10^3$~cm\eee) and $T_1=T/(10$~K) and where
we have also listed the free-fall time for the gas in the box for completeness.
For $\nbht\sim T_1 \sim1$, the simulation approximately
satisfies the observed linewidth-size relation (Solomon et al. 1987; Heyer \& Brunt 2004).
For the remainder of this paper, all results will
be given assuming the fiducial scaling values $\nbh = 1100$~cm$^{-3}$ and $T$=10 K, 
which are appropriate for the Perseus Molecular Cloud (Paper II),
but they may be adjusted to different conditions using equations (2)-(5).
In terms of the Bonnor-Ebert mass,
\begin{equation}
M_{\rm BE}= {{1.182 c^3}\over{(G^3\bar\rho)}^{1/2}}=4.71\frac{T_1^{3/2}}{\nbht^{1/2}}~~~M_\odot, 
\end{equation}
the simulation has a mass of $184 M_{\rm BE}$.
If the Jeans mass is defined as $M_{\rm J} =\rho L_J^3$, where
$L_J=(\pi c_s^2/G\bar\rho)^{1/2}$ is the Jeans length, then
$M_J = (\pi^{3/2} / 1.18) M_{\rm BE}=18.9 T_1^{3/2}/\nbht^{1/2}~M_\odot$.

Our turbulent periodic box study is comprised of two stages. The first stage simulates 
the large scale isothermal environment of a turbulent molecular cloud with self-gravity. In this ``low resolution" stage, we only add enough AMR levels to resolve the shape and structure of collapsing clumps and cores.
This first stage has two parts. First,
to obtain the initial turbulent spectrum, we
turn off self-gravity and use the method described 
in MacLow (1999), in which velocity perturbations are applied to an initially
constant density field. These perturbations correspond to a 
Gaussian random field with flat power spectrum in the range $3 \le k \le 4$ where 
$k$ is the wavenumber normalized to $k_{\rm phys} L / 2 \pi$. 
At the end of two cloud crossing times, the turbulence follows a Burgers $P(k) \propto k^{-2}$ power 
spectrum as expected for hydrodynamic systems of supersonic shocks. 
For the second part, we
turn on gravity and follow the
subsequent gravitational collapse for two scenarios. 
It should be noted that some workers (e.g. Bate et al. 2003) do not allow self-consistent turbulent
density fluctuations to build up before turning on gravity. Any choice of initialization
for a turbulent, self-gravitating cloud is necessarily approximate, but in our view it
is preferable to have self-consistent density and velocity fluctuations in the initial
conditions.
In the simulation that we will 
refer to with the letter D (driven), we 
continue turbulent driving to maintain virial equilibrium, while in the other, noted with U (undriven),
we halt the energy injection and allow the turbulence to decay. 

In the second stage, we select a few 
emerging cores for further study in each turbulent box, and we follow their
fragmentation and evolution into protostellar systems at high resolution using a barotropic equation of state (EOS).
We add additional grid refinement to the regions we select, which continue to  evolve 
within the low resolution context of the box. This method capitalizes on the AMR 
methodology to achieve a high resolution study of the development and 
properties of protostellar cores with realistic initial and boundary conditions. Following all the cores
over a free-fall time with AMR rather than a subset 
to the maximum resolution would require more than a million CPU hours 
on 1.5 GHz processors. 
In contrast, our stage two approach with AMR requires $\sim$50,000 CPU hours per high resolution box.

In the first stage, it is reasonable to assume that the low density gas in the cluster is isothermal
and scale-free, 
reflecting the efficient radiative cooling of the gas. However,
as the gas compresses and becomes optically thicker there is a critical density
at which the radiation is trapped.  Ideally, we would directly solve for the radiation
transfer using an appropriate opacity model to accurately determine the gas 
temperature at these high densities. However, even approximations such as
the Eddington and diffusion approximations do not sufficiently economize the
equations of radiation transfer such that they are affordable over the 
resolution and timescales necessary for this calculation. Consequently, we adopt
a bartotropic equation of state to emulate 
the effect of radiation transfer. 
The gas pressure is given by
\begin{equation}
P = \rho c_{\rm s} ^ 2 +  \left({{\rho} \over {\rho_{\rm c}}}\right)^ {\gamma} \rho_{\rm c} c_{\rm s}^2, 
\end{equation}
where  $c_{\rm s} = ({k_{\rm B} T }/{ \mu})^{1/2}$ is the sound speed, $\gamma=5/3$, the average molecular weight $\mu =2.33m_{\rm H}$, and the stiffening density $\rho_{\rm c}$ is given by ${{\rho_{\rm c}}/ {\rho_{\rm 0}}} = 2.8 \times 10^{8}$. The value of $\mu$ reflects an assumed gas composition of $n_{\rm He} = 0.1n_{\rm H}$.
The value of the stiffening density determines the transition from isothermal to adiabatic regimes. It introduces a characteristic scale into the previously scale-free isothermal conditions.  The isothermal scaling relations above remain valid as long as the ratio of stiffening density to the average box density is presumed to be constant. We chose a value of the stiffening density, $\rho_{\rm c}=2\times 10^{-13}$ g cm$^{-3}$, to agree with the $\rho(T)$ relation calculated by Masunaga et al. (1999), who perform a
full angle-dependent radiation hydrodynamic simulation of a spherically symmetric collapsing cloud core. Unfortunately, we sacrifice some accuracy in using the barotropic approximation in lieu of radiative transfer, since an EOS assumes that gas temperature is a single valued function of density. Simulations have shown that gas temperature in calculations using radiation transfer vs. a barotropic EOS can differ by a factor of several and potentially produce different fragmentation (Boss et al. 2000; Whitehouse \& Bate 2006).

The low resolution initial stage uses a $128^3$ base grid with 4 levels of factors of 2
in grid refinement, giving an effective resolution of $2048^3$. 
The high resolution core study has 9 levels of 
refinement for an effective resolution of $65,536^3$ such that the smallest cell 
size corresponds to $\sim$10 AU for our fiducial values. 

\section { BOUND CLUMP PROPERTIES IN THE LOW-RESOLUTION TURBULENT BOX}

\subsection{Clump Definition}

At the end of a free-fall time,  $t_{\rm ff} = ( {3 \pi} / {32 G \rho_{\rm 0}})^{1/2}$, with gravity we analyze the core properties and compare the driven and decaying turbulent results. At this time, the large scale driven turbulent simulation has 32 sink particles with $14.2 \%$ of the total mass accreted. The decaying turbulence simulation has 20 sink particles containing $13.6 \%$ of the mass. Because the sink particles mark collapsing cores rather than individual protostars at this stage, these percentages should be viewed as an upper limit to the actual star formation rate. Nonetheless, these numbers are not too much larger than the the prediction of a $7\%$ star formation rate per free fall time given by 
Krumholz \& McKee (2005) for our assumed conditions and neglecting outflows. 
In the undriven simulation, the turbulence decays significantly in 1$t_{\rm ff}$ and no new sink particles are formed after $ \sim 0.75 t_{\rm ff}$.  
Without continued driving, there is insufficient energy to create the large
scale compressions responsible for seeding new cores.
After significant turbulent support is lost, the cloud deviates from virial equilibrium and the 
gas falls onto existing over-densities rather than forming new cores. 

In presenting the results from the low resolution simulations,  we restrict ourselves to the analysis of objects that can best be described as ``star-forming bound clumps" (see McKee \& Ostriker 2007), which are generally gravitationally bound but may form several systems of stars. In the following sections, we will adopt the terminology ``core" to refer to the bound condensations out of which a single protostar (i.e. sink particle) or small multiplicity protostellar system forms.  Hence, we do not apply a Clumpfind algorithm, as described by Williams et al. (1994), which also captures unbound and transient over-densities.
Instead, we define a bound core as a sink particle with envelope satisfying
four criteria. First, the density in the included cells must exceed the average shock
compressions, i.e., $\rho \ge \rho_{\rm 0} \cdot {\cal M_{\rm 1D}}^2$, which also ensure a single peak for each core. Second, the total mass in the core must be greater than the local Bonner-Ebert mass, signifying that the core will collapse. 
Each cell, i,  forming a core must also be individually gravitationally bound to it such that
$ |E_{\rm KE}^i| < |E_{\rm PE}^i|$.
Finally, the cells included must lie inside a virial radius, $R$, such that $\alpha_{\rm vir} \le 2$, where
the 1-D velocity dispersion, $\sigma$, is given by the sum of the turbulent and thermal components of the gas velocity:  $\sigma^2 = \sigma_{\rm NT}^2 + c_{\rm s}^2$. We vary the density cutoff by a factor of two and find that changes in the data fits remain within 1-sigma error. Thus, our results are not overly sensitive to our core definition. The larger of these cores may eventually form a cluster of stars and may best be described as star-forming clumps. The smaller cores will likely
%may actually be ``cores" that 
form only a single protostellar system. At the low resolution stage of analysis it is difficult to predict the outcome, and so the line between higher mass star-forming clumps and lower mass cores is ill-defined.

Note that in our methodology, the presence of a sink particle does not {\it guarantee} the eventual formation of a protostar, only that the Jeans condition has been exceeded at some time during the simulation. In each simulation, there are a few sink particles that do not posses envelopes satisfying these criteria. However, all cores included in our analysis are defined to be 
gravitationally bound, collapsing objects rather than transient overdensities
in the flow and hence are predisposed to develop protostellar systems. 

\subsection{Clump Properties}

\begin{figure}
%	\epsscale{.50}
%	\plotone{f1.eps} %rho_cutoff=rhoave*4.9^2
	\includegraphics[scale=0.5]{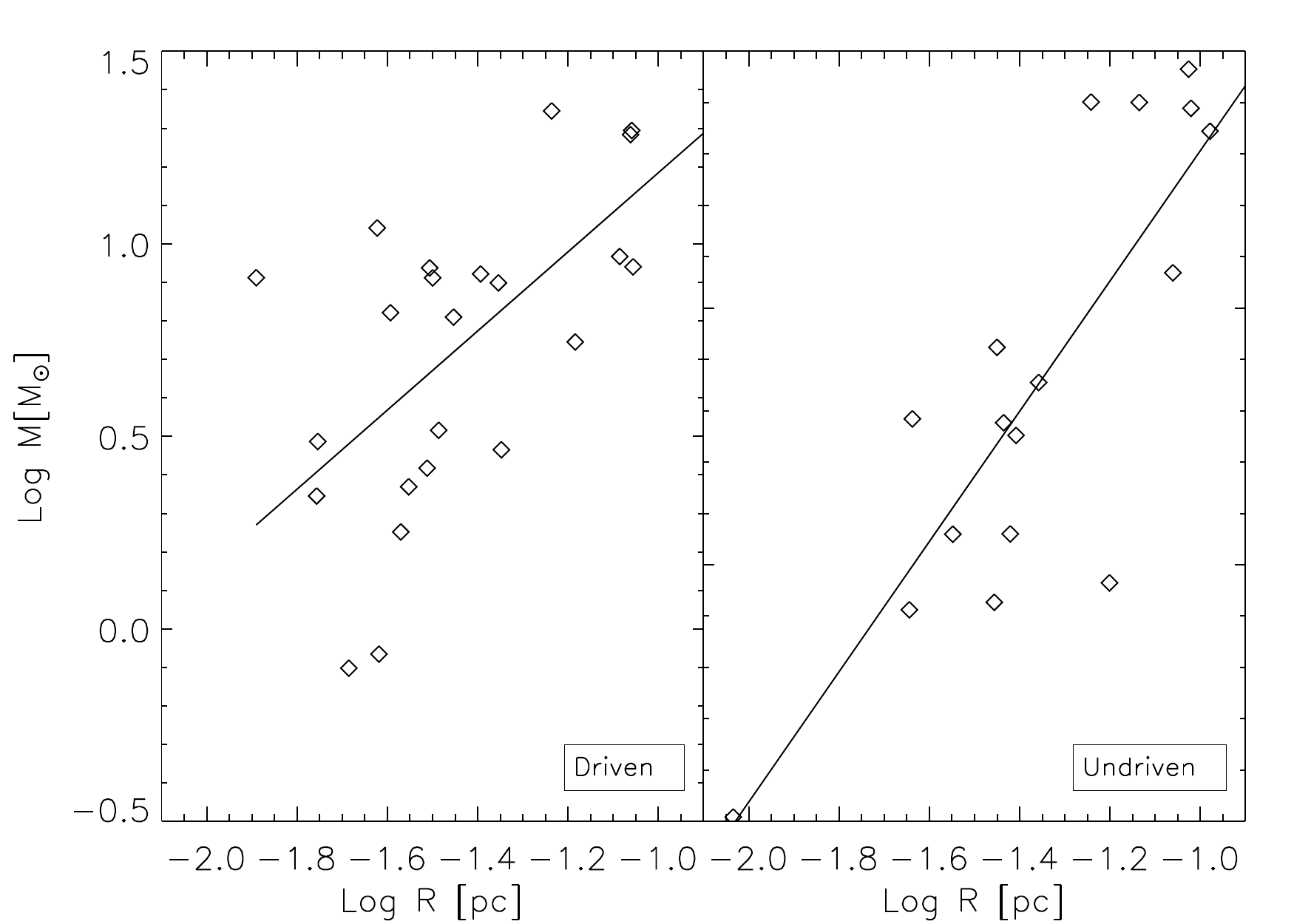}	
	\figcaption { The figure shows the log of the core masses as a function of log size ($R=\sqrt{a b}$) for the driven (left) and decaying (right) boxes at 1 $t_{\rm ff}$.  The slopes have fits of 1.03$\pm$0.26 and 1.27$\pm$0.19, respectively.
         \label{mofr}}	
\end {figure}

\begin{figure}
%	\epsscale{.50}
%	\plotone{f2.eps}%rho_cutoff=rhoave*4.9^2
	\includegraphics[scale=0.5]{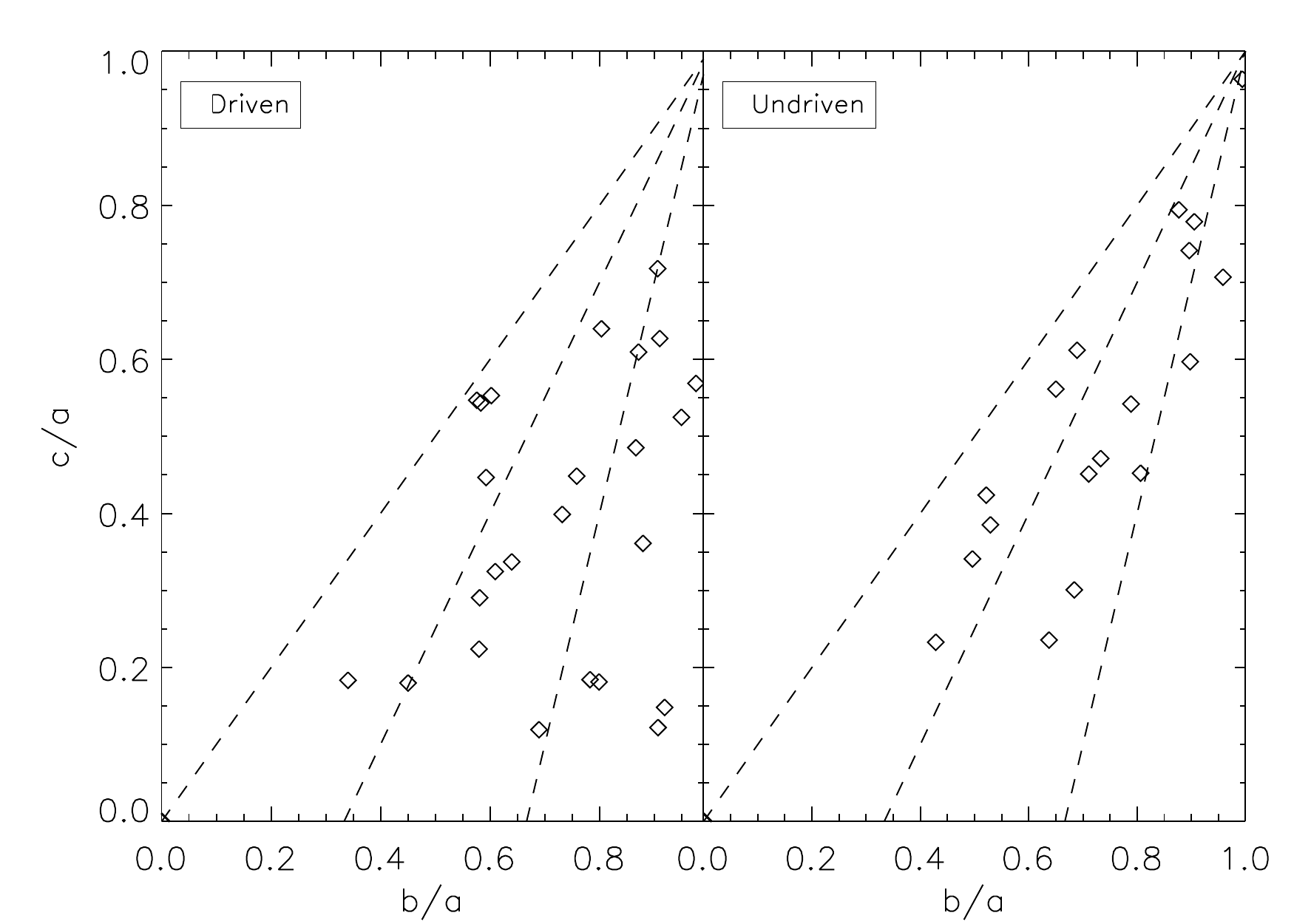}
	\figcaption {The figure shows the core aspect ratios for the driven (left) and decaying (right) boxes at 1 $t_{\rm ff}$. The median aspect ratios for each case are (b/a, c/a) = (0.76, 0.40) and (b/a, c/a) = (0.73,0.54), respectively.
         \label{shape}}	
\end {figure}

\begin{figure}
%	\plotone{f3.eps}%rho_cutoff=rhoave*4.9^2
	\includegraphics[scale=0.5]{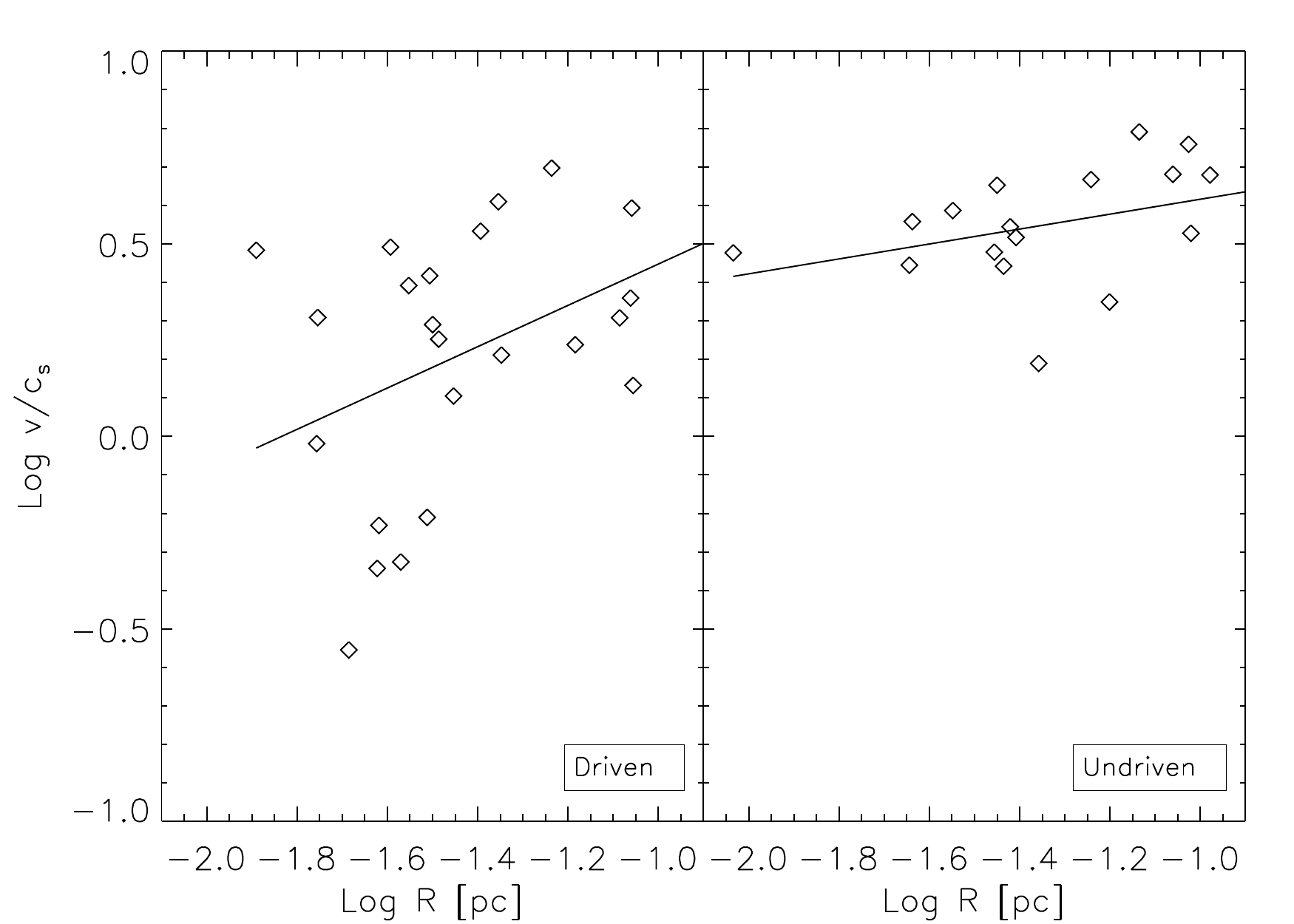}
	\figcaption {The figure shows the log core velocity dispersions as a function of log size ($R=\sqrt{a b}$) for the driven (left) and decaying (right) boxes at 1 $t_{\rm ff}$. The slopes have fits of 0.54$\pm$0.25 and 0.19$\pm$0.11, respectively.
         \label{vofr}}	
\end {figure}

\begin{figure}

	\includegraphics[scale=0.5]{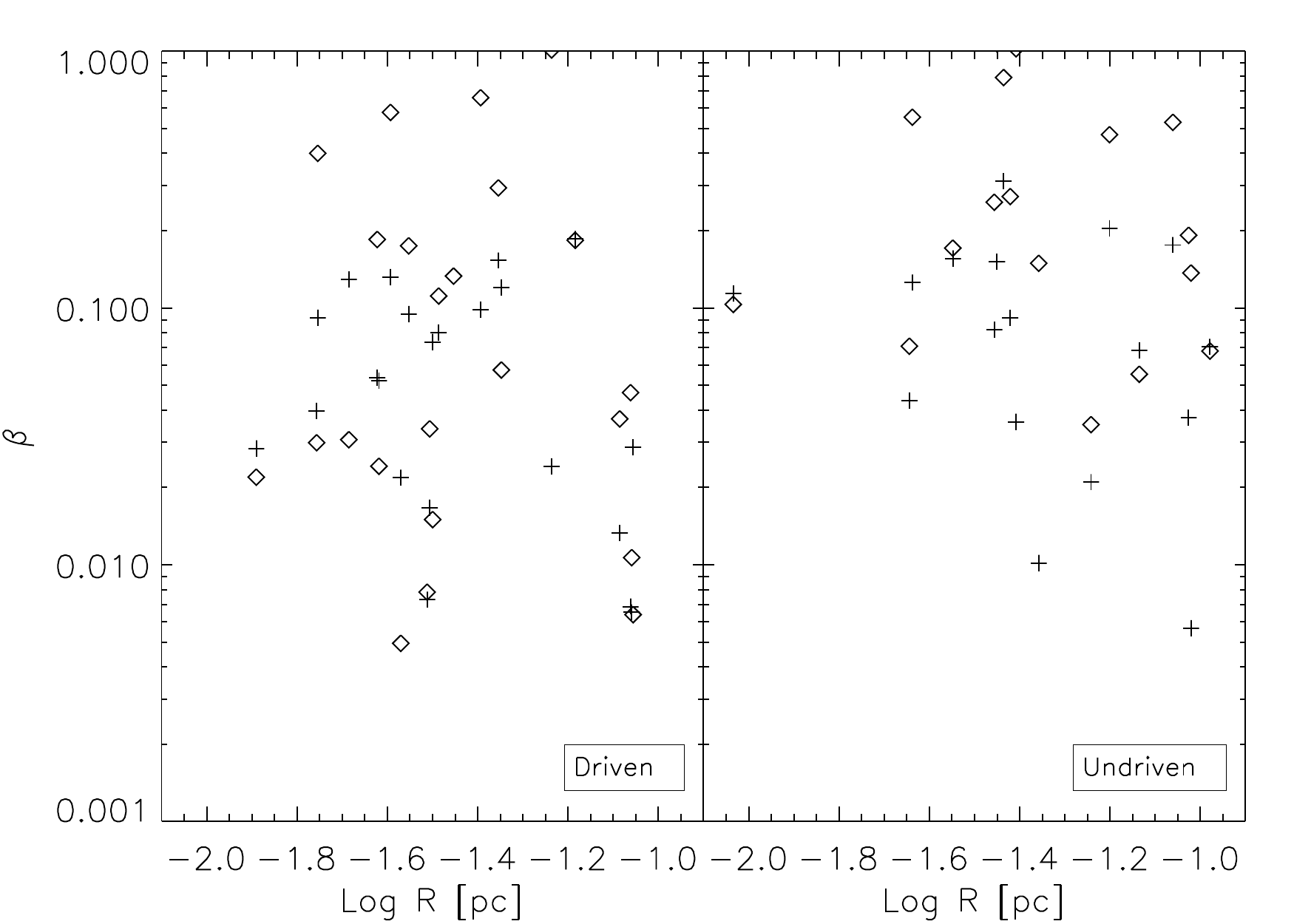}
	\figcaption {The figure shows the rotational parameter, $\beta$, as a function of size ($R=\sqrt{a b}$) for the driven (left) and decaying (right) boxes at 1 $t_{\rm ff}$. The crosses give the 2-D projected value, while the diamonds give the 3-D value. For run D, the median $\beta$ values are 0.05 (crosses) and 0.05 (diamonds). For run U, the median $\beta$ values are 0.08 (crosses), 0.19 (diamonds).
         \label {betaofr}}	
\end {figure}

There are a number of core physical properties that are comparable to
observations, and we investigate these here at 1 $t_{\rm ff}$. 
Figures \ref{mofr} - \ref{jofr} show the bound core data plotted with best fit lines. We exclude objects from the fit that have $R =\sqrt{\rm ab} \le 4\Delta x $, where $a$ and $b$ are the lengths of the major and minor axes. 

%\placefigure{mofr}

 \subsubsection{ Density Profiles}

As plotted in Figure \ref{mofr}, we find that compared to cores in run U, the cores in run D have a slightly flatter trend of  $M(R) \propto R$, consistent with Bonnor-Ebert spheres, which are characterized by $\rho(r) \propto r^{-2.0}$. In run U, the cores have profiles that are closer to a free-fall profile, where $\rho(r) \propto r^{-1.5}$.  Cores that are supported or collapsing slowly will tend to resemble pressure-confined isothermal spheres (Kirk et al 2005, Di Francesco et al. 2007) as in run D, where turbulence is providing more external pressure support. In run U, where the turbulence has decreased significantly, cores tend quickly to infall and collapse as unbound gas becomes gravitationally attracted to the largest overdensities. However, the slopes of the cores in the two simulations are within 1-sigma error due to significant scatter, so that the trends are not significantly different.
 
\subsubsection{ Shape}

As shown in Figure \ref{shape}, both distributions of bound cores have similar morphologies and tend to be mainly tri-axial. It is thought that in the presence of magnetic fields, which we do not include, cores will flatten along the field lines (Basu \& Ciolek 2004). However, ideal MHD simulations by Li et al. (2004) also find that their cores are mostly prolate and triaxial\footnotemark.  In any event, the difficulty of deprojecting observed cores makes the true shape distribution ambiguous. Run D has median major and minor aspect ratios of $b/a$ =0.76 and $c/a$=0.40,  while the decaying cores have median aspect ratios of  $b/a$=0.73 and $c/a$=0.54. The net medians of the shape distributions 0.58 (D) and 0.52 (U) are similar to those observed for different star-forming regions which fall in the range 0.50-0.67 (Jijina et al. 1999).
\footnotetext{Li et al. 2004 and other references use the word  ``core'' to refer to their bound over-densities. For consistency, we continue to use our own definition of cores and cores (see $\cal x$ 3.1). }

\subsubsection{ Velocity Dispersion}

In Figure \ref{vofr}, we plot the velocity dispersion as a function of core size for comparison against Larson's (1981) linewidth-size relation. For low-mass star forming regions, $\sigma_{NT} \propto R^{0.5}$ with some sensitivity to core sizes and clustering (Jijina et al. 1999).  We find exponents of 0.54 (D) and 0.19 (U). The slope of run D is within the range of observed slopes for low-mass regions. Although the scatter in our data appears large, our $\chi^2$ fit slope error is comparable to the range of fit errors ($\pm 0.1 - \pm 0.19$) that Jijina et al. report. Plume et al (2000) observed massive cores with a completely flat slope, and indeed, the cores in run U are more massive
 with a mean mass of 12 \msun versus 8 \msun for the driven,
  but not significantly so (see \S 3.2.6).
 A Kolmogorov-Smirnov (KS)
 test of the distributions of velocity dispersions indicates definitively that the populations are quite dissimilar at the $>> 99\%$ level.
 The difference in slope between the two simulations is possibly due to crowding in the decaying turbulent case caused by insufficient global turbulent support against gravitational attraction.
Jijina et al. (1999) showed that clustered objects have a significantly flatter linewidth-size relation slope. Note, that the magnitudes of the velocity dispersions in run U, although flatter, are higher, which is consistent with quickly collapsing rather than turbulently supported cores.

\subsubsection{Rotation}

Typically, rotational energy makes up only a small fraction of the core gravitational energy. The rotational 
parameter $\beta$ is defined as the ratio of the rotational kinetic energy to the gravitational
potential energy. For a uniform density sphere this can be written:

\begin{equation}
{\beta_{\rm rot}} ={ {1\over 3} {{\Omega^2R^3}\over{GM}} }
\end{equation} 

Observationally, $\Omega_{\rm pos}=\Omega_x^2+ \Omega_y^2$ is the angular velocity projected in the plane of the sky, such that $\beta_{\rm rot, obs}={2\over 3} \beta_{\rm rot}$. 
Goodman et al. (1993), studying a selection of dense cores in NH$_3$, find that $\beta_{\rm rot, obs}$ is roughly 
constant as a function of size and find values of $ 2 \times 10^{-3}  <  \beta_{\rm rot, obs} < 1.4$ with median $\beta_{\rm rot} \sim 0.02$.  Observations of dense cores using N$_2$H$^+$, which primarily traces $n>10^5$cm$^{-3}$, quiescent gas gives similar median value of $\beta_{\rm rot, obs} \sim 0.01$ (Caselli et al. 2002). 
For the purpose of comparison, we evaluate $\beta_{\rm rot}$ in two ways. First, we follow the convention of the observers and evaluate $\beta_{\rm rot, obs}$ by assuming that the cores are projected constant density spheres. 
Second we sum over all the 3-D data to calculate $E_{\rm rot}/E_{\rm grav}$. For a singular isothermal sphere, $E_{\rm rot}/E_{\rm grav} = {1\over 3} \beta_{\rm rot}$.

Figure \ref{betaofr} confirms
that $\beta_{\rm rot}$ for both runs is independent of the core size, and there is fairly large scatter. 
The total range of $\beta_{\rm rot}$ values for observation and simulation is roughly the same. We find a range of 0.0005 $ < \beta_{\rm rot, obs} < $ 0.2 for the driven case and 0.006 $ < \beta_{\rm rot, obs} < $ 0.3 for the decaying case.
However, overall our values are a factor of 2 to 4
higher than those found by Goodman et al. 
Run D has a lower median $\beta_{\rm rot, obs} \sim 0.05$, while run U has very few low $\beta_{\rm rot, obs} $ cores and so has a median $\beta_{\rm rot, obs} \sim 0.08$. 
When we use the complete gas properties to calculate $E_{\rm rot}/E_{\rm grav}$, we find median values of 0.05 and 0.19 for the D and U cores, respectively. Jappsen \& Klessen (2005) 
perform gravoturbulent driven simulations of cores and find a median $ E_{\rm rot}/E_{\rm grav} \sim 0.05$, in agreement with our result.
The higher $\beta_{\rm rot,obs}$ values measured in the cores in the undriven simulation may be a side effect of the smaller turbulent support: Since the U cores are moving more slowly,
they may more easily accrete gas from farther away, which has higher angular momentum.
(We thank the referee for this comment.)  Although a KS test verifies that the two 
$\beta_{\rm rot, obs}$ populations are distinct, neither is a good match for
observation since both have
median values that are higher than observed.

One possible explanation for the factor of 3-5 difference between simulation and observation is that magnetic fields play a significant role in decreasing core rotation. A number of recent simulations of isolated rotating magnetized cloud cores have shown that magnetic braking is an efficient means of outward angular momentum transport (Hosking \& Whitworth 2004; Machida et al. 2004; Machida et al. 2006; Bannerjee \& Pudriz 2006).  
The oblate cores formed in the ideal MHD simulation of Li et al. (2004) show a median $\beta_{\rm rot, 3D}$ similar to ours (Li, private communication), however, all their cores are supercritical by an order of magnitude.

Another possibility to account for the difference in median $\beta$ is that observers typically investigate isolated cores, which are easier to distinguish and analyze but tend to be less turbulent. However, our study specifically concerns bound cores forming in a turbulent cluster. Using Larson's laws,  $\beta_{\rm rot} \propto v_{\rm rot}^2/ (GM/R) \propto R/(GM/R) \propto 1/\Sigma \simeq $const.  However, there is large scatter and a few exceptions of clouds with non-constant column density, $\Sigma$, such that measurements of $\beta_{rot}$ could be sensitive to differences in column density in various MCs. 

\begin{figure}
%	\plotone{f5.eps}%rho_cutoff=rhoave*4.9^2
	\includegraphics[scale=0.5]{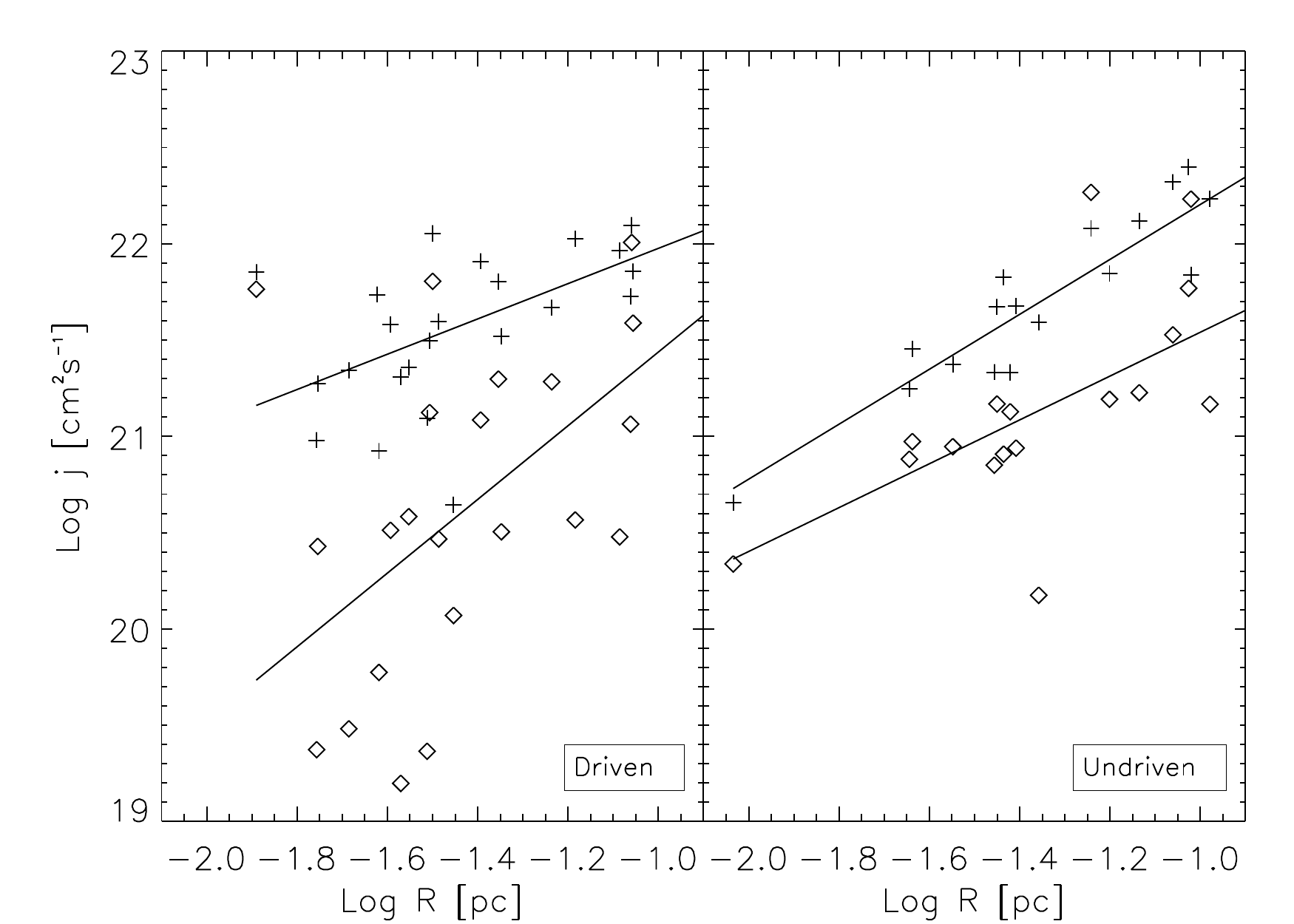}
	\figcaption {The figure shows the log of the core specific angular momentum as a function of log size ($R=\sqrt{a b}$) for the driven (left) and decaying (right) boxes at 1 $t_{\rm ff}$.  The crosses give the 2D projected value, while the diamonds give the 3D value. For run D, the slopes have fits of 1.91$\pm$0.65 (diamond), 1.14$\pm$0.31 (cross). For run U, the slopes have fits of 1.14$\pm$0.35 (diamond), 1.50$\pm$0.23 (cross).
        	 \label{jofr}}
\end {figure}

\subsubsection{ Angular Momentum}

There is also a substantial difference between the specific angular momentum in the two cases as illustrated in Figure \ref{jofr}. We plot both the 3D 
total specific angular momentum
of the cores, 
which is obtained by directly summing the angular momentum of the individual cells comprising a clump, and the 2D specific angular momentum, by totaling the projected momentum along a line of sight.
In run D, the specific angular momentum fits, $j_{\rm 2D}(R) \propto R^{1.1}$ and $j_{\rm 3D}(R) \propto R^{1.9}$, bracket the expected $j(R) \propto R^{1.5}$ based upon the linewidth $\delta v \propto R^{1/2}$ and assumption of virial balance (Goodman et al. 1993;
this argument suggests the same value for both the 2D and 3D cases).
The specific angular momentum fits in run U are more similar but still a little flat, $j_{\rm 2D}(R) \propto R^{1.5}$ and $j_{\rm 3D}(R) \propto R^{1.1}$.
Because the decaying cores are less turbulent, they will be inclined to have less variation
of angular momentum than their driven counterparts.
The cores in run D overshoot the expected 
relationship for $j_{\rm 3D}$,
 while the decaying cores undershoot by a similar amount.
In either case, we expect that the simulated angular is affected by
the absence of braking effects from magnetic fields, which we do not 
include in these simulations. 
Nonetheless, 
we find that the measured range of $j \sim 10^{21}-10^{22}$ cm$^2$ s$^{-1}$ to be consistent with observational estimates and the 2D angular momentum estimates to be statistically similar to one another,
but flatter than the measured $j_{\rm 2D} \propto R^{1.6 \pm 0.2}$
(Goodman et al. 1993; 
see Fig. \ref{jofr}).

\subsubsection{Core Mass Function}

Measurements of the core mass function (CMF) show that its shape strongly resembles the stellar initial mass function (IMF) (Lada et al. 2006). The high mass end, in particular, seems to share a similar power law index. As a key characteristic of star formation, the core and star mass functions for driven and undriven turbulence has been extensively numerically studied. Ballesteros-Paredes et al. (2006) and Padoan et al. (2007) find a mass function of the form ${d\cal N}/d{\rm log}(m) \propto m^{-1.3}$ for cores in driven hydrodynamic turbulence, even without the presence of self-gravity. Klessen (2001) finds that both driven turbulence with $1 \le k \le 2 $ and undriven turbulence produce a core spectrum with a similar slope to that of the measured IMF. A number of isothermal SPH simulations of decaying
 turbulence have shown agreement with the observed IMF 
 despite different initial turbulent conditions in which a turbulent velocity spectrum is initialized on a constant density field and then allowed to decay in the presence of self-gravity (e.g. Klessen \& Burkert 2001; Bate el al. 2002; Bonnell et al. 2003; Tilley \& Pudritz 2004; Bonnell et al. 2006). In this method, the turbulence does not reach a steady state and the simulated cloud is not virialized as observed.

\begin{figure}
%	\plotone{f6.eps}
	\includegraphics[scale=0.5]{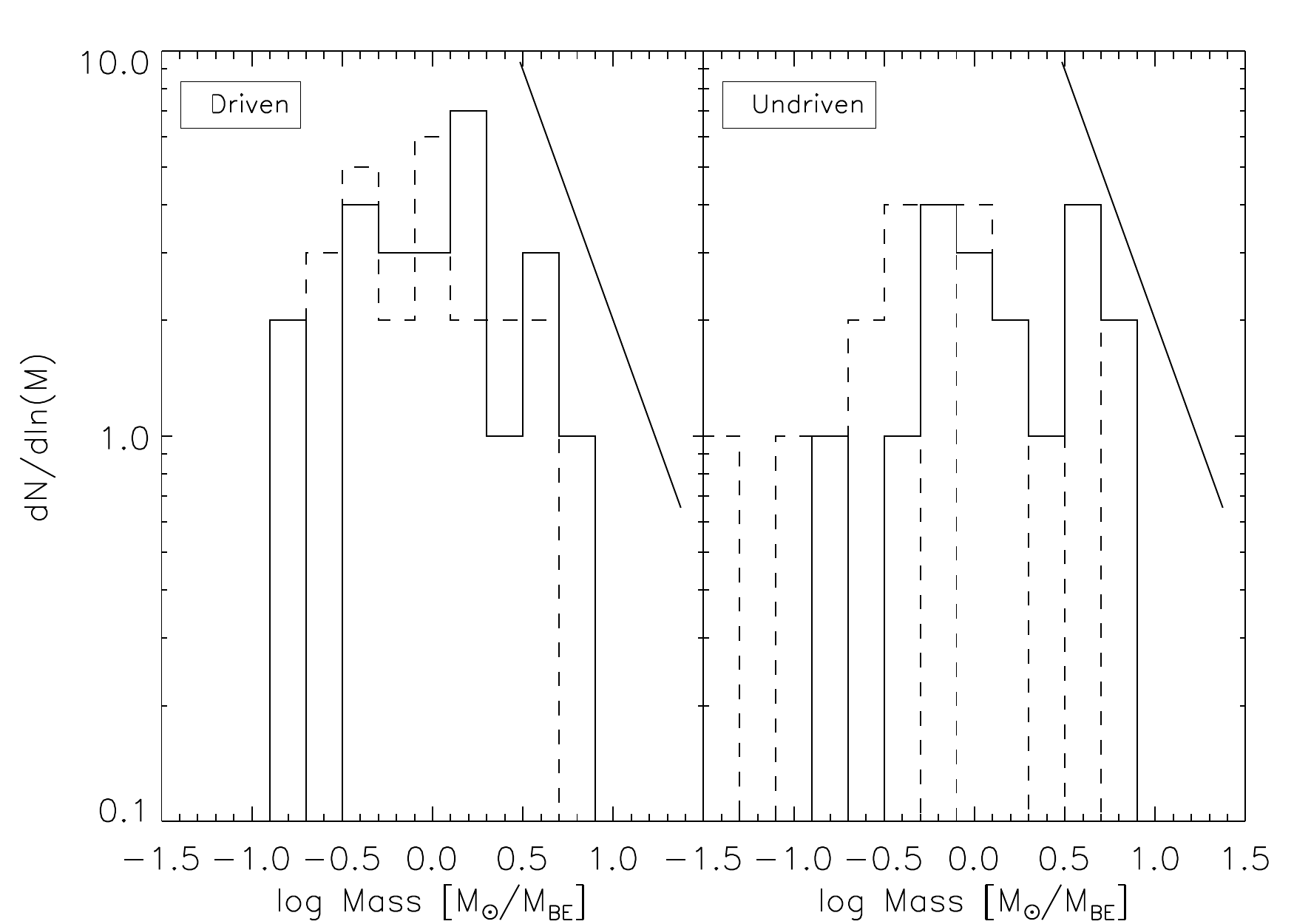}
	\figcaption {The figure shows the sink (dashed line) and core (solid line) mass distributions for the driven (left) and decaying (right) runs at 1 $t_{\rm ff}$. The straight time has a slope of -1.3. 
         \label {Mhist}}	
\end {figure}

For the purpose of comparison, we plot the CMF for the simulations D and U at 1 $t_{\rm ff}$ in Figure \ref{Mhist}. The two runs produce 30 and 19 bound cores, respectively. 
Unfortunately, the statistics at the high mass end are too small to be able to rule out either distribution on the basis of agreement with the Salpeter slope. Although the agreement looks better for the driven cores, we find that the mass distributions are in fact statistically similar according to a KS test.

Overall, the simulations have statistically different distributions of angular momentum, rotational parameter, and velocity dispersion. The decline of turbulent compressions in the undriven run appears to make some significant changes and causing fewer new condensations to be formed. As turbulent pressure support is lost, the contracting gas instead falls onto existing cores resulting in less turbulent, more quickly rotating cores than in the driven case. 
However, it is not possible at this time to say which approach corresponds more closely with observation.

\section { PROTOSTELLAR CORES AT HIGH RESOLUTION}

\subsection { Overview}

\begin{deluxetable*}{lrrrrcrrrrrrr} 
\tablecaption{Low resolution core properties
for each case when the sink particle has 0.1 $\msun$.}
\tablehead{
\colhead{} 	& \colhead{U1a/U1b} & \colhead{D2} & \colhead{U2} & \colhead{D3} & \colhead{U3}}   
\startdata
%Core Mass (\msun) 	& 13.25	&	6.25 		  	 & 5.34	& 1.27 &	3.47\\  %T=20,nH2=2.5d3
Core Mass (\msun) & 10.71   &    5.05  &    4.32 &     1.59 & % T=10, nH=1.116d3
       2.80 \\
%$R_{max}$ (pc)	&  0.276	&      0.14  			 & 0.12	& 0.02 &	0.05\\
 $L_{\rm max}$ (pc) & 0.45   &   0.23   &   0.19  &   0.03 &
     0.08 \\
 Shape  			& 1:0.28:0.05	 & 1: 0.37:0.09 	 &1:0.66:0.26 & 1:0.69:0.60 & 1:0.78:0.24\\
%$v_{rms}$ (km/s)	&0.60	& 0.46			 & 0.51 	& 0.47 &	0.64\\   
$v_{\rm rms}$ (km/s)	&  0.42   &  0.32 &    0.36  &   0.33  &   0.45 \\
 $n_{\rm ave}(10^{5}cm^{-3})$ & 1.44	& 1.08    			 & 1.01 &  4.06	& 1.25	\\
 $\alpha_{\rm vir}$ 		&	1.5	& 1.12			 &1.90 & 1.06	& 2.24	\\
 $\beta_{\rm rot}$ 			&	0.052	& 0.025		& 0.011  & 0.028	 & 0.047	\\
 %$ t_{ff} (10^{4} yr)$  &	3.9	&    4.5  			 &4.7     & 2.3 &		   2.6       \\
 $ t_{\rm ff} (10^{4} yr)$ &  8.9 &      10.3   &    10.7  &     5.3  &  5.9 \\
 ${\cal M}_{\rm 1D}$ \tablenotemark{a}   	&	2.4	&	4.9	& 3.9		 & 4.9 &		3.0	\\
\enddata
\tablenotetext{a} { $m _{\rm 1D}$ is the velocity dispersion of the entire box, which is fixed at 4.9 for the driven cases.}
\label{table1}
\end{deluxetable*}

In this section, we present our computational results for the evolution of the protostellar systems contained in a few selected cores using 5 or 6
additional levels of refinement. We accomplish our study by inserting a 
refinement box around the core of interest before a sink particle is introduced on level 4
so that cells inside the box
continue to higher densities and refine according to the Jeans criterion, while cells in
the remainder of the simulation refine only to a maximum level of 4 as before. The lengths of the high resolution boxes are typically 0.25-0.5 pc depending on the size of the enclosed clump. As a result, the boxes contain a region $\sim$200-2000 times volumetrically smaller than the simulation domain. The total initial mass in the boxes ranges from 4-12 $\msun$, which easily encompasses the bound core and all collapsing regions associated with it. In this way, we can perform a high resolution study of selected collapsing cores with realistic initial conditions and consistent boundary conditions taken from the surrounding lower resolution grids computationally cheaply and efficiently. In each 
portion of the box, sink particles are introduced when the corresponding maximum
refinement level is reached. At high resolution, each sink particle represents a single  ``protostellar core." 
Thus, we are able to follow the clump
fragmentation at high resolution, without the need to re-run the entire calculation
at that resolution.

We chose six cores for further study. In cases U1a and U1b, we test for convergence by 
following the same cores at two different resolutions. In cases D2 and U2, we choose an early 
collapsing object that is present in both the driven and decaying simulations to highlight the differences between the calculations. The cores U1, D2, and U2 have initial bound masses greater than 2.5 $\msun$. We also study two smaller driven and undriven cores, D3 and U3.  As a result, we observe the effect of turbulent support on protostellar system development. 
The physical properties of the selected cores are given in Table \ref{table1}. Table \ref{table2} gives the three dimensionless quantities relating to the box surrounding each core: $\cal M$, $\alpha_{vir}$, and the self-gravity parameter, 
\begin{equation}
 \mu\equiv \frac{M}{c_s^3/(G^{3/2}\bar\rho^{1/2})},
\end{equation}
where $\alpha_{\rm vir}$ can then be written as
\begin{equation}
\alpha_{\rm vir}=\frac 56\left(\frac{{\cal M}^2}{\mu^{2/3}}\right).
\end{equation}
These three parameters characterize the amount of turbulence in the core vicinity, the degree of self-gravitization of the gas, and the extent to which balance is achieved between the two.
Table \ref{table2} indicates that all the small boxes are subsonic and thus the influence of gravity is dominating the gas in the regions around the cores.

Note that when we cease driving in run U, the turbulent cascade continues and the turbulent decay rate is determined by the Mach number and the domain size as described by Mac Low (1999). 
At any given time, the effect of the decay on the cores forming in the high resolution subdomain depends upon the amount of turbulent decay in the large box. The 1-D velocity dispersion in Table \ref{table1} is an indicator of the change in turbulent energy when the core of interest is collapsing.

\begin{deluxetable}{lrrrrcrrrrrrr} 
\tablecaption{Turbulent box properties for the whole domain and the small boxes containing the cores.}
% pltdtb31851 ; pltdtb42522 ; pltvpbsfs1925; pltdtb31890; pltvpb42143; pltdtb42042
\tablehead{
\colhead{} & \colhead{D/U at t=0} & \colhead{U1a/U1b} & \colhead{D2} & \colhead{U2} & \colhead{D3} & \colhead{U3}}   
\startdata
%Note: These are independent of normalization
$ {\cal M}_{\rm 3D}$	 		 & 8.37          &  0.71	& 0.80		& 0.72   &  0.38	&0.74 	\\
$ \mu_{\rm box}$  	 		&  206.82	 & 19.03	& 10.74              & 12.41 &  2.6 	& 4.59	\\
 $\alpha_{\rm vir}$ 		          & 1.67      &0.06	& 0.11		& 0.08   & 0.06	&0.16 	\\
\enddata
\label{table2}
\tablecomments{The values for the small boxes are determined using the length $L_{\rm small}$=0.25 pc.} 
\end{deluxetable}

\subsection {Convergence Study}
 
\begin{figure}
%\plotone{f7c.eps}
\includegraphics[scale=0.5]{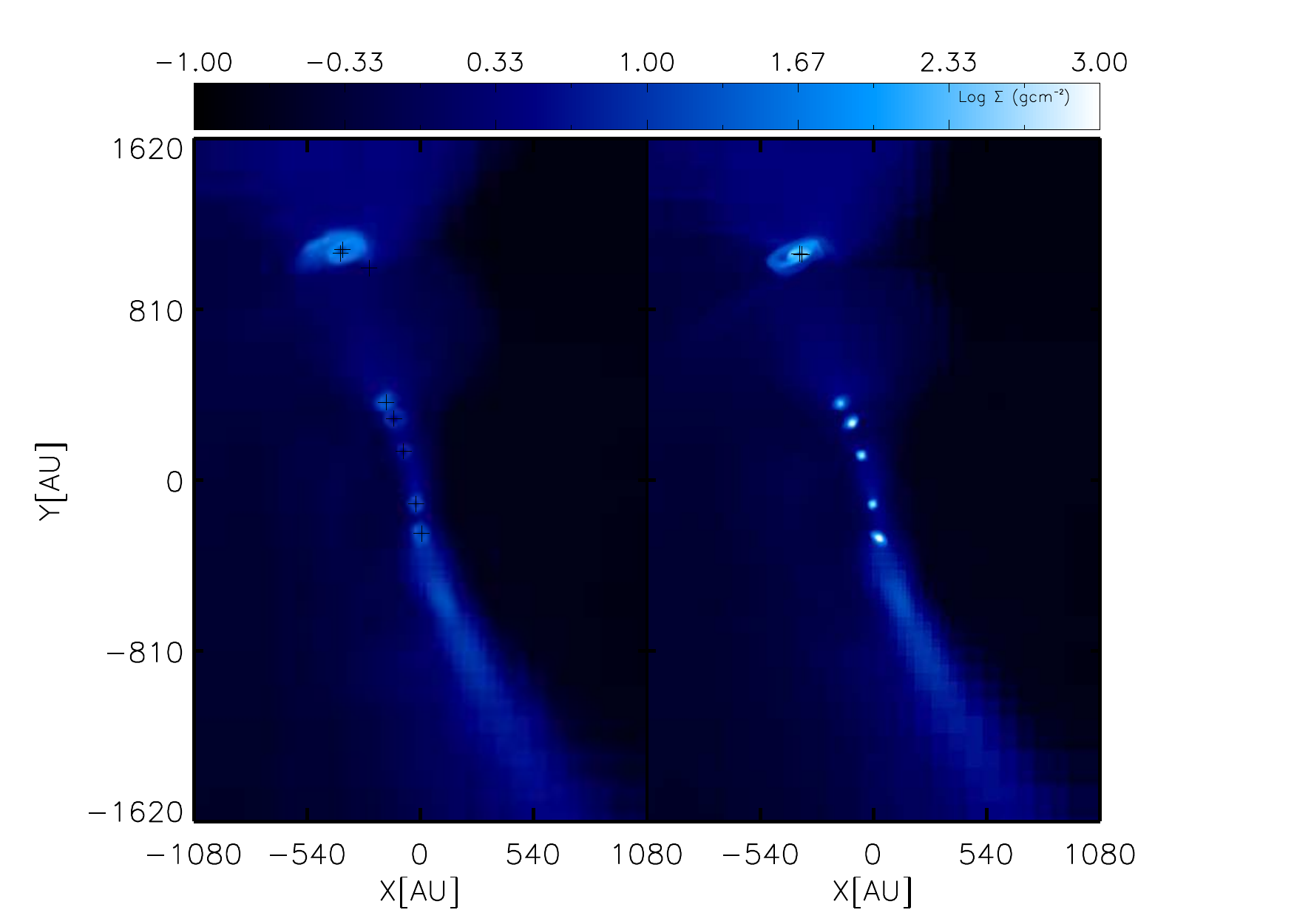}
\figcaption{The figure shows the log column density of a core in the decaying turbulence simulation U1b (left) and U1a (right) with resolution of 10 AU and 5 AU 16 kyr after the formation of the first sink particle. 
\label{logcol3kyr}}
\end{figure}

\begin{figure}
%\plotone{f8c.eps}
\includegraphics[scale=0.5]{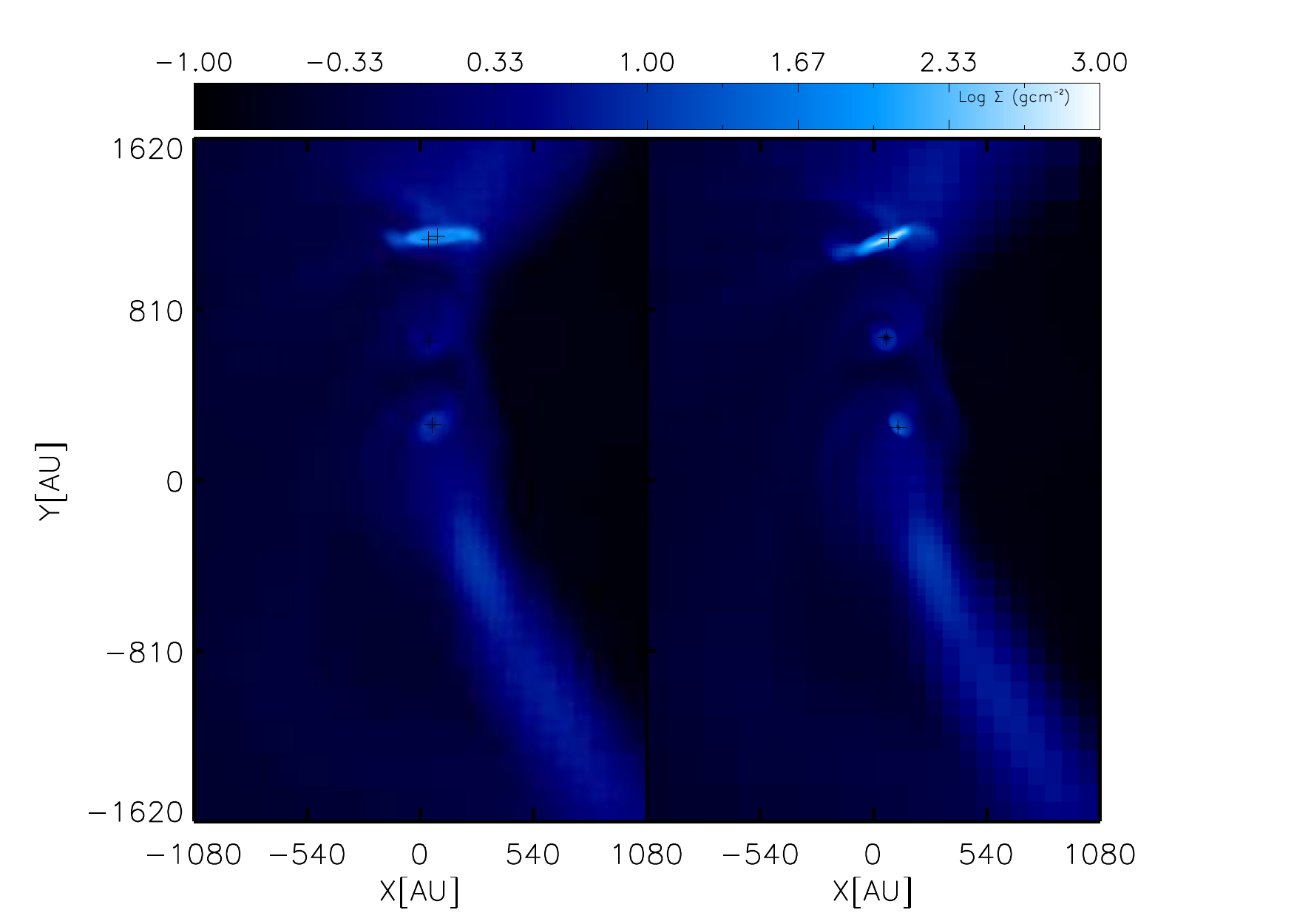}
\figcaption{The figure shows the log column density of a core in the decaying turbulence simulation U1b (left) and U1a (right) 23 kyr after the formation of the first sink particle. 
\label{logcol7kyr}}
\end{figure}

\begin{figure}
%\plotone{f9c.eps}
\includegraphics[scale=0.5]{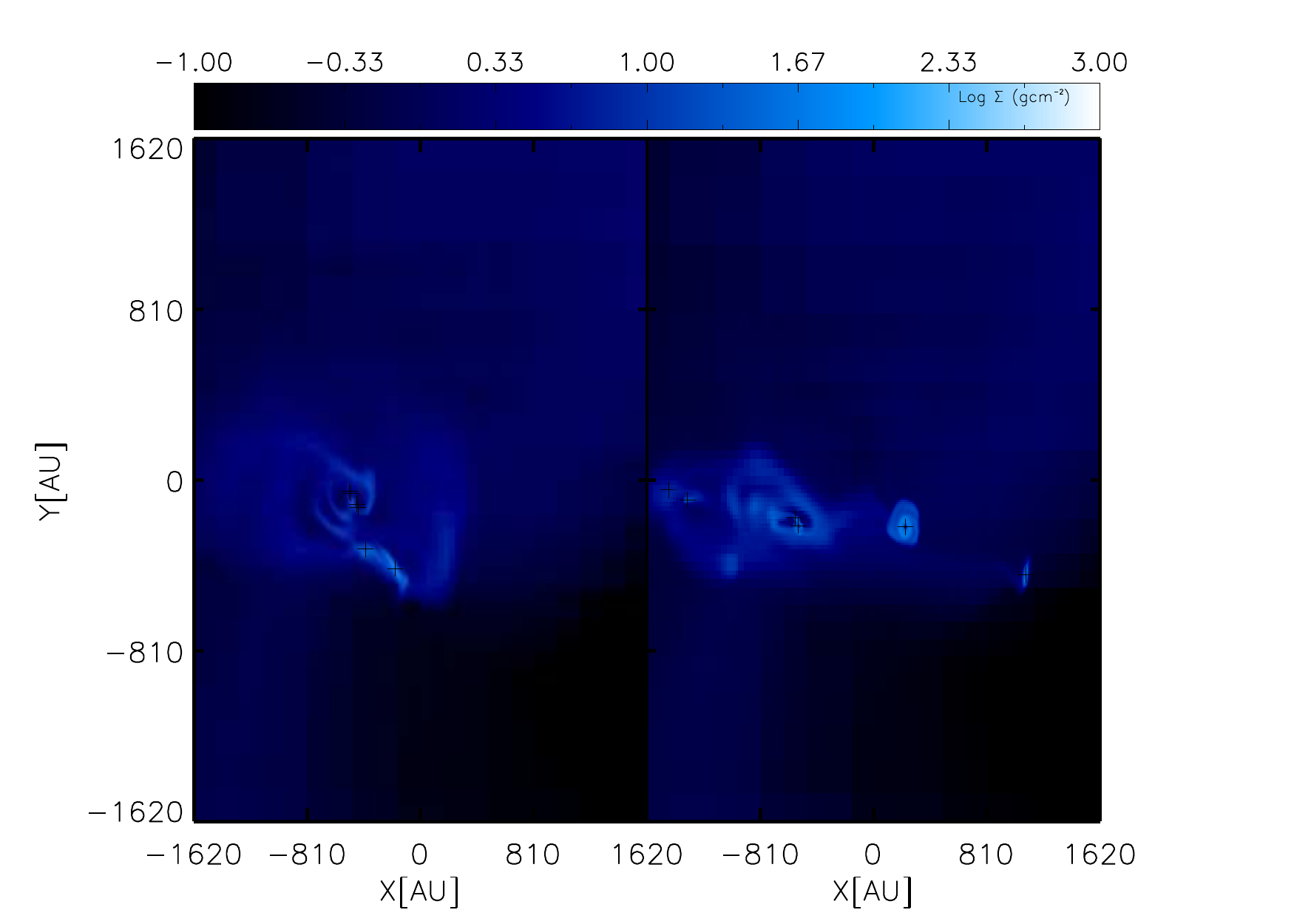}
\figcaption{The figure shows the log column density of a core in the decaying turbulence simulation U1b (left) and U1a (right) 53 kyr after the formation of the first sink particle. 
\label{logcol23kyr}}
\end{figure}

Before embarking on further analysis, it is important to show that the results at the calculation resolution are suitably converged.  In particular, it is necessary to show not only that there is no artificial fragmentation but that the number of fragments is constant with increasing resolution.  For our convergence study we consider a box in the decaying turbulence run, U1, which encloses a long filament that collapses to form a number of small over-dense fragments along its length. We run this calculation with 9 (U1b) and 10 (U1a)  levels of refinement, which corresponds to a minimum cell size of $\sim$10 and 5 AU, respectively. Figures \ref{logcol3kyr}, \ref{logcol7kyr} and \ref {logcol23kyr} show the two simulations at 16 kyr, 23 kyr, and 53 kyr, respectively, after the formation of the first sink particle. Tables \ref{table3} and \ref{table4} give the sink particle masses and the fragment masses at these times.  We define the fragments as discrete cores of bound gas with density greater than $2 \times 10^{-16}$ g cm$^{-3}$.

We find that both resolutions produce the same number of collapsing fragments and yield a similar collection of sink particles. Most of the fragment masses for the two resolutions differ by at most a few percent, while sink particle masses may differ by 50\%. At a particular instant in time, discrepancies between the number of sink particles in the two runs can occur due to several factors. First, the addition of extra levels allows the higher resolution simulation to collapse for a longer time without exceeding the Jeans criterion. Thus, a sink particle ultimately forms in both cases at the same location but at slightly different times.  Another possibility is that a sink particle forms in both simulations at similar locations, but in one it mergers with a larger neighbor.  A final possibility is that the region that collapses in the higher resolution becomes thermally supported before a sink is formed. In all these cases the gas physics can be quite similar but the introduction of sink particle can differ due to small details. 
For example, at 3 kyr the low resolution simulation has formed sink particles in each filamentary fragment with condensation masses ranging from $1.5\times 10^{-2}-8\times 10^{-2} \msun$ (see Table \ref{table4}), while the higher resolution run has not reached sufficient density for any sink particles to form.  This rather odd filamentary structure is created and confined by the ram pressure of intersecting shocks. As a result it forms somewhat smaller bound clouds than the minimum Bonnor-Ebert mass associated with the local pressure ($P \simeq 3\times 10^6$ dy cm$^{-2}$)  at $\rho \simeq 10^{-14}$ g cm$^{-3}$. The smallest sink particles formed in the filament later merge as shown in Figure \ref{logcol7kyr} when the gas in the filament streams onto the disk-protostar system. 

At later times and for small masses the corresponding sink particle properties differ the most significantly, particularly at the lower mass end as shown in Table \ref{table3}.  Due to the intrinsically chaotic and dynamically unstable nature of three or more body systems, at later times the evolution of the two calculations begins to diverge. This is unsurprising because not only do the calculations have different AMR grid structures, but the particle members of the system are introduced at slightly different times and initial masses. Despite this, the masses and configuration still show reasonable agreement at 80 kyr.

\begin{deluxetable} {r r r r r }
\tablewidth{0pt}
\tablecaption{ Masses of the stars in $\msun$ for decaying simulations at two different 
resolutions at two different times after the formation of the first sink particle. }
\tablehead{
\colhead{ $\Delta$ t} & \multicolumn{2}{c}{23 kyr} & \multicolumn{2}{c} {53 kyr} \\
\colhead{Resolution}& \colhead{ 5 AU } & {10 AU }   & \colhead{ 5 AU }  & \colhead{10 AU  }}
\startdata
 &    0.834 &    0.705  &   1.224   &   0.925  \\
 &     0.000  &   0.216  &   0.000  &   0.369 \\
 &    0.264   &  0.262   &  0.571  &   0.455 \\
 &     0.171   &  0.175  &   0.762  &   0.768 \\
 &    0.000   &  0.036  &   0.106  &   0.141 \\
 &   \nodata   &  \nodata  &   0.061  &   0.036 \\
  &  \nodata   &  \nodata  &   0.128   &  0.180 \\
\enddata
\label{table3}
\tablecomments{The subscripts 10 (U1a) and 9 (U1b) represent 
the number of AMR levels. The sink particle absence in the second row of the high resolution column is due to an early merger (m < 0.1$\msun$ with the neighbor listed in the first row.}
\end{deluxetable}

\begin{deluxetable} {r r r r r r r}
\tablewidth{0pt}
\tablecaption { Core gas mass ($\msun$) including embedded sinks for the decaying simulations at two different 
resolutions at three different times.}
\tablehead{
\colhead{$\Delta$ t} & \multicolumn{2}{c}{16 kyr } &  \multicolumn{2}{c}{23 kyr } & \multicolumn{2}{c}{53 kyr  } \\
\colhead{Resolution}& \colhead{ 5 AU } & {10 AU }   & \colhead{ 5 AU }  & \colhead{10 AU  }  & \colhead{ 5 AU }  & \colhead{10 AU  }  }
\startdata
 &    0.741 &    0.758  &   0.997   &  0.997   &  1.907   &  1.905 \\
  &   0.124  &   0.124   &  0.280  &   0.283 &    0.827 &    0.806\\
 &    0.077   &  0.077 &    0.177  &   0.180 &    0.136   &  0.191\\
  &   0.040  &   0.037  &   \nodata  &   \nodata &    \nodata &   \nodata\\
  &   0.035  &   0.033  &   \nodata  &  \nodata  & \nodata &  \nodata\\
  &   0.291 &    0.259  &  \nodata   & \nodata  & \nodata & \nodata\\
 &\nodata & \nodata &  \nodata  &   \nodata &    0.106  &   0.036\\
\enddata
\label{table4}
\tablecomments{The subscripts 10 (U1a) and 9 (U1b) represent 
the number of AMR levels. The minimum density of the gas is $\rho =  2 \times 10^{-16}$ g cm$^{-3}$. The '...' represent cores that have merged with others and cannot be individually distinguished.} 
\end{deluxetable}

\subsection {Influence of Turbulence on Stellar Properties}

Interstellar turbulence undoubtedly has a substantial effect on cloud lifetimes 
and core creation, however, its relationship with core fragmentation and evolution
is less certain.  The level of turbulence in cores is partially dependent on how much
mass and energy the envelope exchanges with the surrounding turbulent gas.
In turn, the properties of the parent core influence the rate of protostellar core formation 
and accretion.
If substantial mass continues to fall onto the clump, as in the case of global
contraction, then external flow patterns will impinge upon on the system development, increasing the accretion rate and possibly causing fragmentation.
If however, the core accretes at a relatively low level in the manner of Bondi-Hoyle 
accretion in a turbulent medium then the core will accrete much less 
over time (Krumholz et al. 2006; Krumholz et al. 2005). Protostars forming in such a core limit to the Bondi-Hoyle accretion 
rate as the high density gas is depleted.
 
 In cases D2 and D3 we continue turbulent driving to maintain virial equilibrium. 

 To avoid directly adding artificial perturbations that may
 affect the core development or seed new fragmentation, we do not apply
 any velocity perturbations to the high resolution regions inside the refinement box. 
Thus, the turbulence cascades into the highly-refined box from the outside in a 
 self-consistent manner.
 In cases U2 and U3, the simulation continues without any turbulent injection. 
 
We find striking differences in the protostellar systems formed in the driven and decaying cores. The most obvious difference between the D and U runs is the difference in the number and mass  
of sink particles formed (Table \ref{table5}).  For example, initially the fragmentation of D2 and U2 is 
similar temporally and spatially, but U2 eventually forms a slightly larger number of objects  particularly at small masses as the level of turbulence in the two simulations diverges. The small D3 core forms a small stable binary system at early times, whereas U3, which is also fairly small, fragments into a number of protostellar members.

\begin{deluxetable}{r r r r}
\tablewidth{0pt}
\tablecaption { Masses of the protostars in both driven and decaying simulations at 260 kyr (the larger cores, D2 and U2) and 130 kyr (the smaller cores, D3 and U3). }
\tablehead{ \colhead{ D2 ($\msun)$}   & \colhead{U2 ($M_{\odot})$}& \colhead{D3 ($M_{\odot})$}& \colhead{$U3 (M_{\odot})$}}  
\startdata  
     1.221  &   1.811  &   0.639  &   0.586 \\
     1.047  &   1.002  &   0.453  &   x 0.552\\
     1.049  &   0.933  &  \nodata   &   0.348\\
     0.490 &  x 0.223  &  \nodata  &x   0.114\\
  x 0.382  & x  0.131  &   \nodata &   0.048\\
     0.329  &   0.059  &   \nodata &  x 0.047\\
     0.281  &  x 0.034  &   \nodata &   \nodata \\ 
     0.207  &   x 0.030  &   \nodata &   \nodata  \\
 \nodata  &   x 0.023  &   \nodata &   \nodata
\enddata
\tablecomments{ The x's represent particles that are ejected from the system by dynamical interactions. The time of first sink particle formation after the onset of gravity for each of the cores is 270, 680, 250, and 660 kyr for D2, D3, U2, and U3, respectively.}
\footnote{One additional BD mass sink particles have been excluded as numerical disk fragmentation from column U2.}
\label{table5}
\end{deluxetable}

Despite the small-number statistics, we are able to compare the IMF
of the protostars to the observed initial mass function via the KS test. 
The KS test determines the probability that a given data set is drawn from a specified statistical 
distribution, in this case, the single star IMF given by Chabrier (2005).
This test is accurate for input sets of 4 or more data points. We achieve a best fit by scaling the masses by an adjustable normalization, $\epsilon = m_*/m_{\rm sink}$, where $m_{\rm sink}$ is the mass of the sink particle. 
Given that this simulation
lacks feedback effects such as outflows and radiation transfer, the sink 
particle masses represent an upper limit and the scaling factor corresponds to an efficiency 
factor of $\epsilon$=0.25-0.75  (Matzner \& McKee 2000).

\begin{figure} 
%\plotone{f10.eps} 
\includegraphics[scale=0.5]{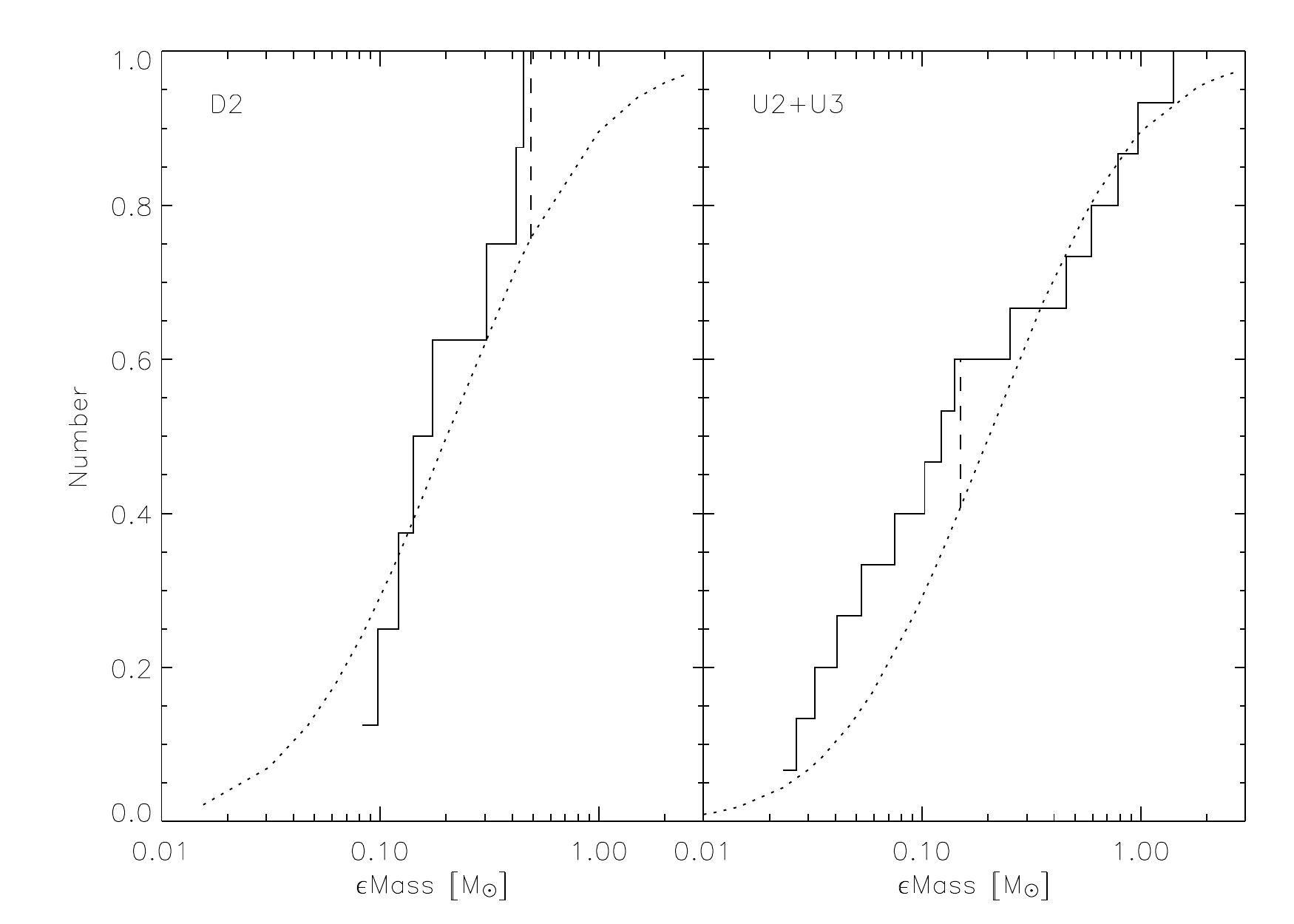}
\figcaption{The figure shows the cumulative distribution function (solid line) at t=0.26Myr for D2 ($\textit{left}$), U2+U3 ($\textit{right}$),where the dotted line is the Chabrier 2005 IMF fit. The dashed vertical line represents the point of largest disagreement. The probability that the data are drawn from the Chabrier IMF is 67\% and 59\%, respectively, where the efficiency scale factors of the simulations are 0.4 and 1.0, respectively.  
\label{IMF_plot}}
\end{figure}

 Figure \ref{IMF_plot} shows the scaled cumulative distribution function (cdf) for the runs D2, U2, and U3, and the cdf of the Chabrier is overlaid for comparison. 
Although all three runs can have a high confidence level of agreement with the measured IMF, the normalization values and the shapes of the distributions are quite different.
For example, the smaller stellar population of D2 has fewer low-mass objects and hence has a smaller efficiency scaling factor of $\sim$ 0.4 with highest likelihood of being drawn from the IMF of 67\%.
Conversely, U2+U3 distribution contains 
collections of low-mass objects and intermediate mass objects, where the largest disagreement
occurs in the middle of the two populations. A scaling factor of 1.0 gives the best agreement of 59\%. 
A scaling factor near unity implies that protostellar mass loss has a negligible  effect
on the final mass of the star, contrary to some theoretical expectations 
(Shu et al 1987; Nakano et al. 1995; Matzner \& McKee 2000).
For the D2 distribution, the largest disagreement occurs at the higher mass end, indicating that if the protostars continue to accrete mass
and no new protostars are formed,
then it is likely that the high probability of  agreement with the IMF will be maintained while the scale factor shifts to a 
lower value. The U2+U3 have a larger scale factor due to the significant number of low mass objects with accretion halted by dynamical ejection. These objects will be unlikely to accrete additional mass and are essentially fixed. For the undriven runs, the largest disagreement occurs in the middle of the distribution, indicating a widening difference between the sub-stellar fixed-mass ejected objects and those that remain in the gas reservoir and continue accreting. Further running time will more likely make the gap wider and agreement worse.

The efficiency scale factor is also dependent upon the normalization we have chosen. The minimum mass that we are able to resolve in these simulations is proportional to the Jeans mass evaluated at the maximum level of refinement. Because the Jeans mass is inversely related to the density, normalizing the results to a density higher than our fiducial value of $n=1100$ cm$^{-3}$ will produce lower mass objects and shift the IMF peak towards lower mass. This will also increase the efficiency factor used to scale the distribution to the universal IMF. 

\begin{figure}
%\plotone{f11.eps}
\includegraphics[scale=0.5]{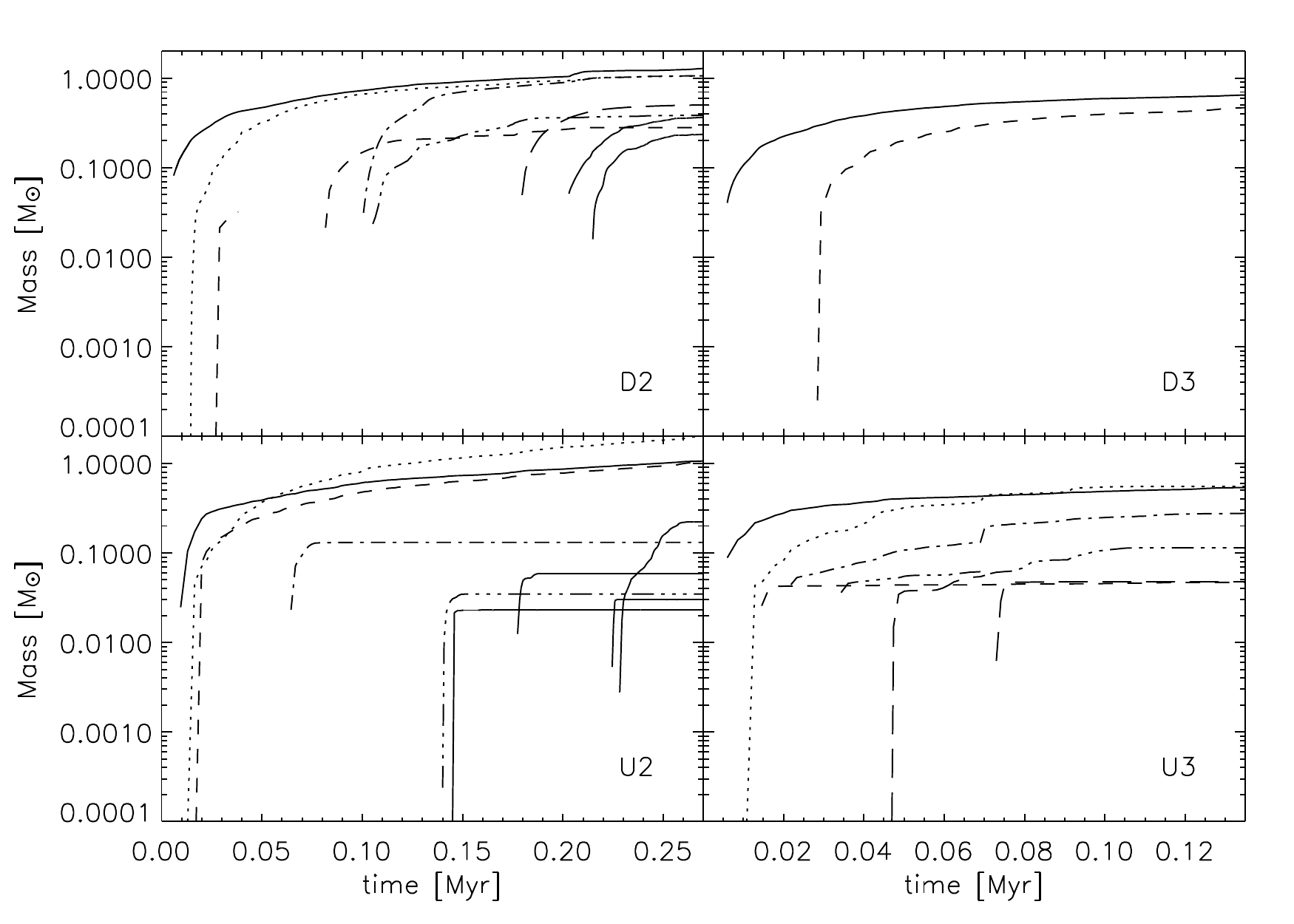}
\figcaption{The figures show the sink particle mass as a function of time for runs D2, D3, U3, and U2 shown clockwise from top left. Each particle is represented by a different style line. 
\label{Mass_plot}}
\end{figure}

Studying the time evolution of the two simulations shows the origin of the different stellar
populations. In simulation D2, the initial collapse and core fragmentation
produces three well separated objects that remain fairly far apart. A few additional objects 
form, but they do not suffer large gravitational interactions with the primaries and
so remain in the high density regions and continue accreting. By contrast, in U2 and U3 the lack of
global turbulent support causes mass to fall onto the early formed protostellar cores resulting in 
contraction of the clump. This causes all the protostars to gravitate towards the core center. As the protostellar proximity increases, the accretion disks interact and 
become gravitationally unstable (see discussion in $\cal x$ 4.6). 
Fragmentation ensues. The stellar systems become increasingly
dynamically unstable with the addition of these small latecomers, which rapidly suffer strong gravitational 
interactions with the larger protostars and are thrown out of the high density reservoir of gas. Their
small envelopes are stripped away, thus truncating the accretion process and effectively fixing their
stellar masses (see Figure \ref{Mass_plot}). This truncation process occurs for approximately half 
of the objects formed in the undriven simulations (see discussion of brown dwarfs in $\cal x$ 4.5)

\subsection{Accretion}

There are two main accretion paradigms. In both models, star formation begins with the outside-in formation
of gravitationally bound cores and their inside-out collapse.  However, the core accretion model proposes that the main protostellar accretion
phase takes place early on and continues until the entire core mass is accreted or expelled (Shu et al 1987; Nakano et al. 1995; Matzner \& McKee 2000), after which
accretion becomes negligible. Thus, the initial core size and subsequent feedback effects limit the mass of the protostars. In contrast, the competitive accretion model proposes that stars begin in a core as wandering, accreting 0.1 $\msun$ seeds, whose final mass is determined by the protostar's location in the clump (Bonnell et al. 1997, 2001). Mass segregation is a common feature of this model, such that the largest mass objects inhabit the region of highest gravitational potential and the smallest objects inhabit the less dense gas, usually having been ejected from the center by gravitational interactions.

In our results, there are two accretion phases. Initially, there is a transient period of high accretion during which the initial infalling gas accretes onto the newly formed sink particle and gas is depleted from the cells inside the accretion region. This phase is model independent and occurs while the newly created sink particle region reaches pressure equilibrium with the surrounding gas. Generally less than 10\% of the accretion occurs during this time. During the second phase, the accretion rate 
approximates the Shu model for core accretion,
\begin{equation}
\dot M_{*} = 
c_s^3 /G,
\end{equation}
(Shu et al. 1987),
although in most cases the accretion rate is gradually declining.
This solution is valid until the rarefaction wave reaches the core,
i.e. when approximately half of the original core mass has been accreted 
(McLaughlin \& Pudritz 1997; McKee \& Tan 2002).
After that, the accretion rate diminishes as the density of the surrounding gas decreases,
but our simulations end before it is possible to determine if this final stage actually occurs.
 
\begin{figure}
%\plotone{f12.eps}
\includegraphics[scale=0.5]{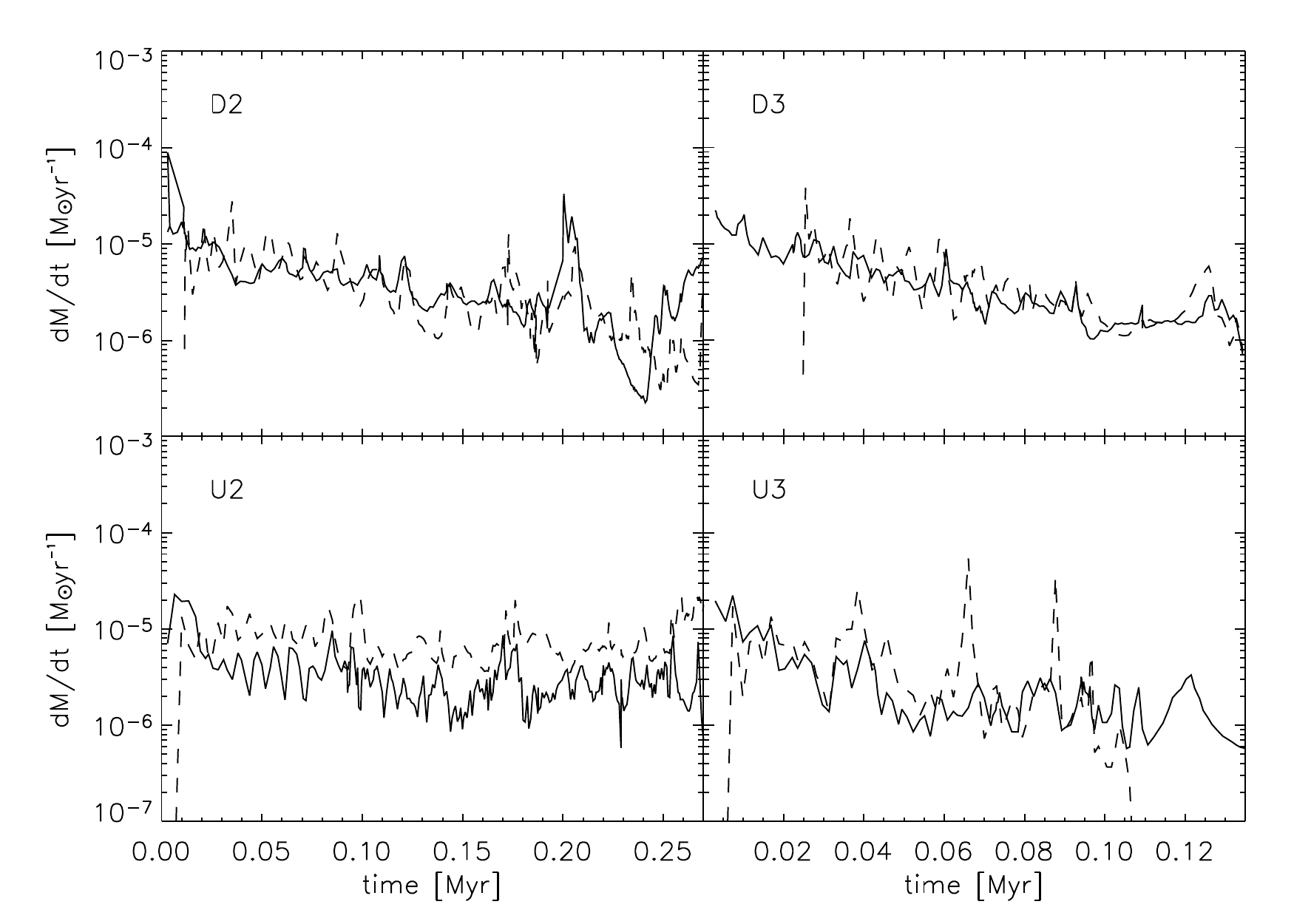}
\figcaption{The figures show the instantaneous sink particle accretion rate as a function of time for runs D2, D3, U3, and U2 shown clockwise from top left. Only the history of the two first forming particles is shown.
\label{Mdot_plot}}
\end{figure}

To illustrate the differences between the protostellar systems, we have plotted the mass as a function of time for all sink particles, the instantaneous mass accretion rate for the first two formed objects, the time-averaged accretion rate, and the total mass in sink particles as a function of time.
The turbulent core accretion and competitive accretion models describe the evolution of the stellar population in cases D2 and U2, respectively. In the former case, objects are mainly formed from core fragmentation with separations larger than 1000 AU. 
The average accretion onto the protostars initially agrees with the Shu model but, modulo fluctuations,  diminishes over time as the core mass depletes (Figure \ref{Mdot_plot}). Meanwhile, the core envelope accretes according to the Bondi-Hoyle model of turbulent accretion (Krumholz et al. 2006).
For driven turbulent environments the overall Mach number will remain sufficiently high such that the core will not gain a substantial amount of mass during the core dynamical time and the main accretion phase of the forming protostars will be limited by this time.  However, in the decaying turbulent case loss of turbulent pressure support potentially causes significant additional mass to accrete onto the core, resulting in a more constant protostellar accretion rate
(Figure \ref{Mdot_plot}, bottom row). However, the differences in the accretion rates of the most massive objects are subtle due to the significant fluctuations.

Perturbations to the accretion disks and clumpiness of the infalling gas cause fairly large variability in the sink particle accretion rate as illustrated in Figure \ref{Mdot_plot}.  However, we do not observe that most of the mass is deposited in short intervals by clumpiness in the disk as noted by Basu et al. (2006), who model 2D axi-symmetric disks with magnetic fields. The absence of this effect in our calculations is most likely due to our Cartesian geometry rather than lack of magnetic fields (Basu, private communication). The $r-\phi$ geometry used by Basu et al. is more suitable for disk treatment and has lower numerical viscosity, which may suppress small scale clumpiness.  

\begin{figure}
%\plotone{f13.eps}
\includegraphics[scale=0.5]{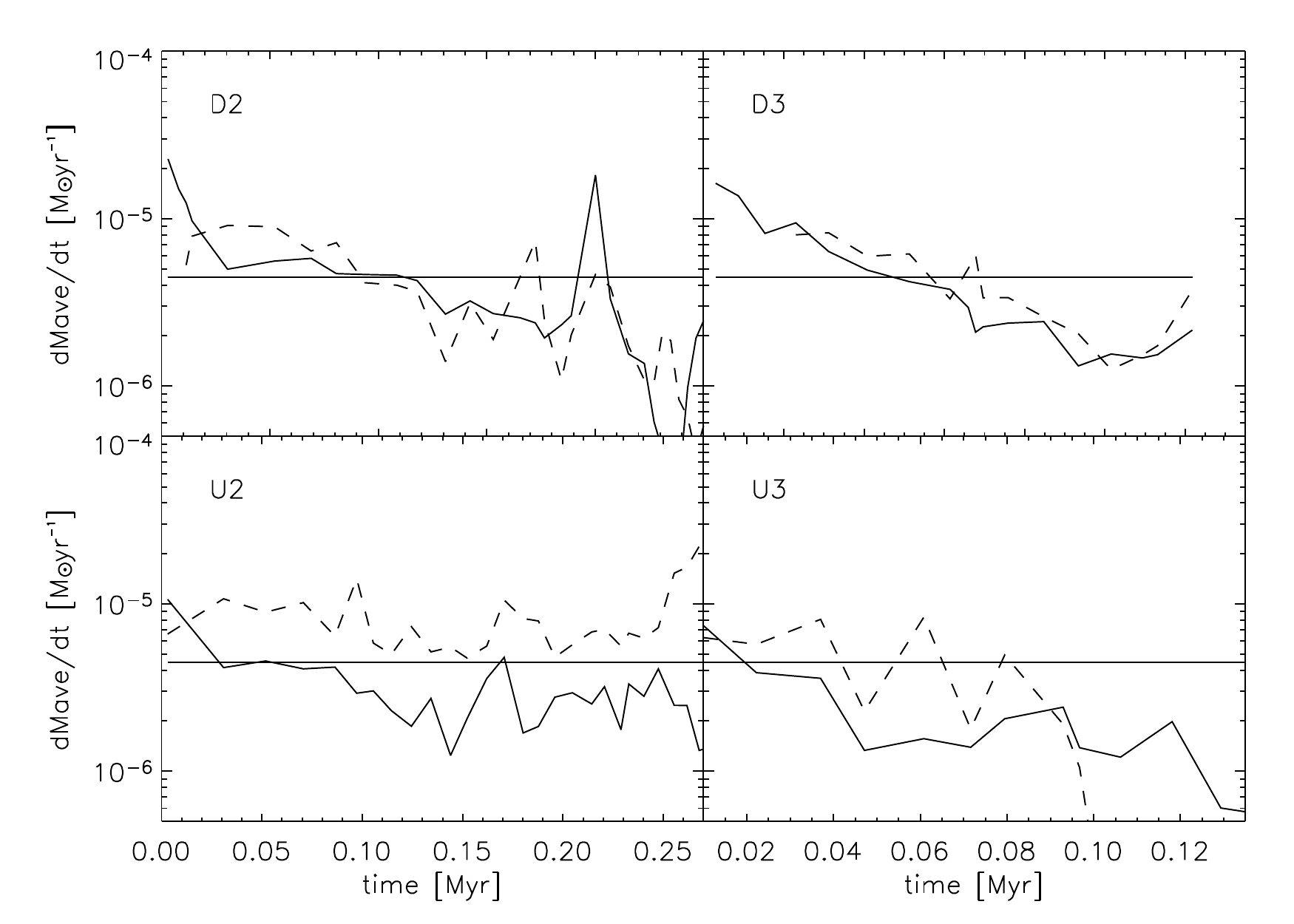}
\figcaption { The figures show the averaged sink particle accretion rate for the first two sink particles as a function of time for runs D2, D3, U3, and U2 shown clockwise from top left. The average is taken over 10 consecutive timesteps, and the solid flat line indicates the value of $c_{\rm s}^3/G$.  
\label{Mave_plot}}		
\end{figure}

\begin{figure}
%\plotone{f14.eps}
\includegraphics[scale=0.5]{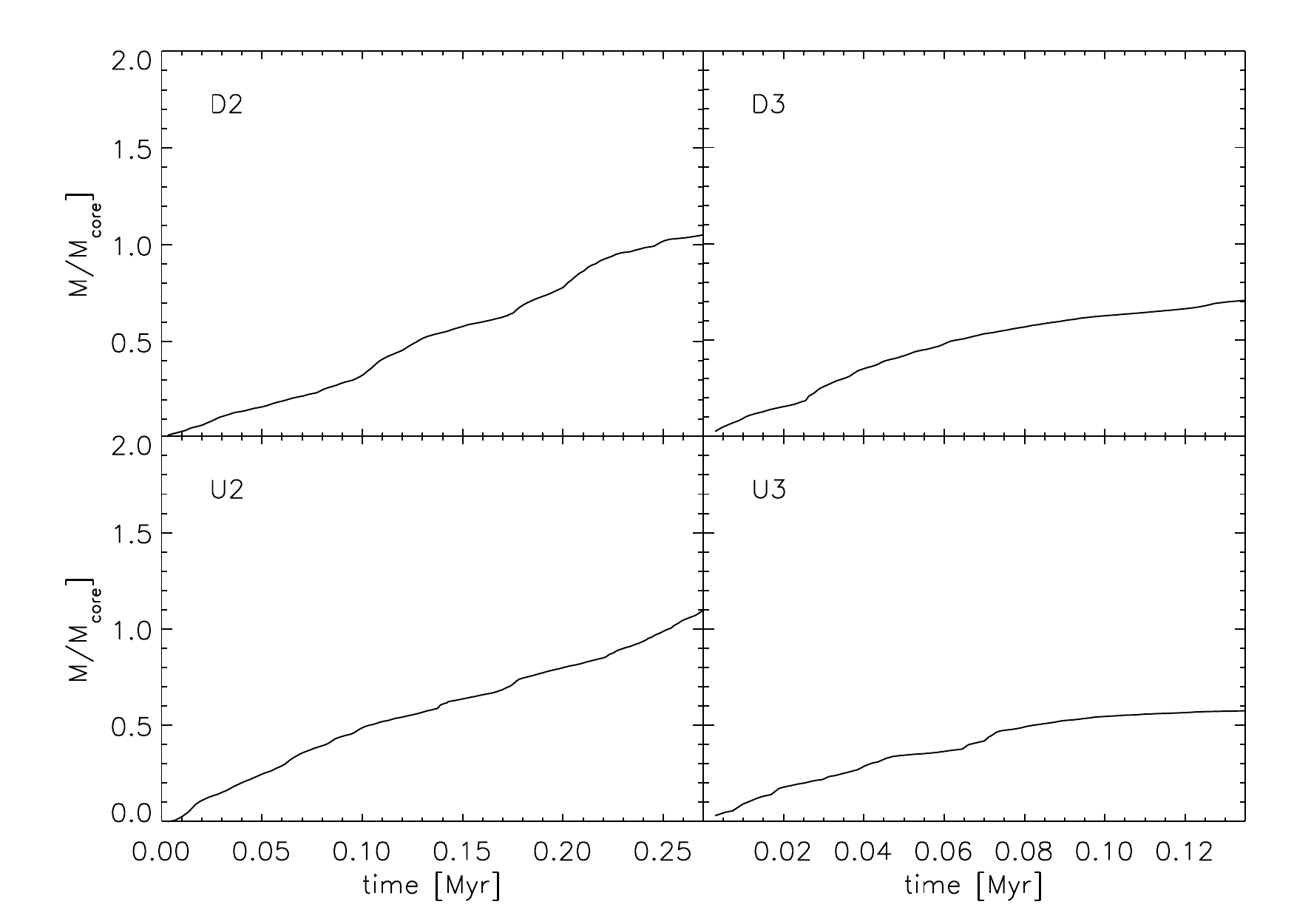}
\figcaption{The figures show the total mass accreted normalized to the initial bound core mass as a function of time for runs D2, D3, U3, and U2 shown clockwise from top left. 
\label{Mtot_plot}}
\end{figure}

Due to differences in core accretion, the two cases produce much different stellar populations. In the driven cases, in which fewer objects form, protostars accrete more smoothly and do not undergo strong dynamical interactions with their neighbors. However, in the decaying cases, the strong infall pushes the protostars to the core center and the large number of nearby objects accreting from the central reservoir of gas causes the smallest objects to be kicked out of the cluster. This can be observed in Figure \ref{Mdot_plot} (U3) as a precipitous drop off in the accretion rate for individual objects or as flatlining of the object mass (Figure \ref{Mave_plot}, bottom row). Differences between the number of objects are caused by turbulent support, which prohibits new mass from infalling onto the clump. High accretion causes fragmentation and drives forming objects to the gravitational center, where they dynamically interact. This close proximity results in object ejection and destabilization of the accretion disks, which leads to new fragmentation. The differences in accretion rate and stellar population between the two cases suggests that the maintenance of turbulence, protostellar accretion, and stellar population are intimately related (Krumholz et al. 2005). In spite of individual accretion fluctuations, the total accretion of the objects from the core is dominated by the largest objects, such that the fraction of the core accreted is relatively smooth over time as shown by Figure \ref{Mtot_plot}.

%SSRO 11/6/07 plot_mdot_all

%SSRO 11/6/07 plot_mdot_all2

\subsection{Brown Dwarfs}

Brown dwarfs (BD) are observed to comprise $\sim$10-30 \% of all luminous objects in star forming regions (Andersen et al. 2006; Luhman et al. 2007). For example, in the Chabrier (2005) IMF, with which we compare, BDs with masses $M_{*} \le 0.08 M_\odot$ comprise $\sim$20\% of the total number of objects. Understanding the population, origins, and connection between planets and hydrogen-burning stars is essential in formulating a successful theory of star formation. Observations remain particularly ambiguous concerning the primary formation mechanism of BDs, sparking many theories. Of these, proposals for BD formation by turbulent fragmentation, ejection, or via disk fragmentation have the most potential for generating BDs in sufficient numbers (Padoan \& Nordlund 2004; Reipurth \& Clarke 2001; Whitworth et al. 2007). Simulations provide an important vehicle for testing these theories, and we discuss the BD population of our simulations in this section.

Our driven turbulence simulations D2 and D3 do not produce any sink particles of final substellar mass, which are primarily created in our simulations by prematurely truncated accretion. However, if BDs are formed via the same mechanism as stars, accrete from a disk, and produce outflows (Luhman et al. 2007) then the same efficiency factor will be valid for scaling all the objects in the simulation. 
The decaying turbulence simulations U2 and U3,  however,  produce a much larger initial number of BDs, 33\%. This agrees with the competitive accretion paradigm:
Bate et al. (2002) find that $\sim$44 \% of the objects that form in their 50$M_{\odot}$ decaying turbulent cloud simulation qualify as BDs, which is much higher than the observed fraction. 
Their calculation is initialized with uniform density and an initial turbulent velocity field, however they do not drive the turbulence, so that the turbulence never achieves a steady relaxed state. Despite this difference, their result is in qualitative agreement with the BD formation mechanism and number fraction of our decaying turbulence runs.

Ideally, we would like to understand the BD population in various star forming regions as a function of their general properties. The turbulent fragmentation model predicts an upper limit on the total mass available for the formation of BDs as a function of the Mach number and average number density (Padoan \& Nordlund 2004). According to their model, the total gas mass $\it{available}$ to make BDs from turbulent compressions is 0.4\% of the total gas mass or 3.7 $\msun$ as a function of our simulation Mach number and density. 
If the SFR per free fall time for the driven and decaying runs is respectively 14.3\% and 13.6\% then the total maximum possible mass in BD due to turbulent fragmentation as a fraction of the actual mass turned into stars is 2.8\% and 2.9\%. For comparison, the fraction of the actual luminous mass turned into BDs according to the Chabrier IMF is $\sim$2\%. Our high resolution protostellar systems have a BD mass fraction of 0.0\% and 3.2\% for D2, and U2 + U3, respectively, using the efficiency factor from Figure \ref{IMF_plot} and including $M_{*} \le 0.08 M_\odot$ in the BD mass total. However, the turbulent fragmentation model gives only the maximum fraction of gas that can be converted to BDs by turbulent compressions and it does not include possible BD formation in disks (Goodwin \& Whitworth, 2007). Fragmentation of disks and dynamical ejection is responsible for all of the BDs in the decaying simulation. Thus, comparison between the decaying turbulence models and turbulent fragmentation prediction is misleading.

The absence of BDs in the driven runs is reasonable if BDs actually form via turbulent fragmentation. In such a process, small low-mass objects form from small low-mass cores. Since we have not chosen any particularly small cores for high resolution study, we would not expect to find many BDs. Thus, scaling to the stellar IMF, which has a peak at $\sim$ 0.2 $\msun$ requires a small efficiency factor.  The core distribution is set by the low resolution turbulent initial conditions. The minimum expected core mass is the Bonnor-Ebert mass evaluated at the maximum turbulent gas density. According to the turbulent fragmentation model, this maximum is set by the probability density function (PDF) of the gas density. The resolution and Mach number of our simulation yield a density PDF  that falls off at $\sim 1.3 \times 10^{-18}$ g cm$^{-3}$ or $M_{\rm BE} \simeq 0.2M_\odot$.  The minimum mass estimated from this density agrees with the minimum sink particle mass at the end of a free fall time at low resolution. Since this mass is well above the maximum BD mass, it also explains the low abundance of low-mass objects at high resolution in the driven simulation. Moreover, this suggests that the driven high-resolution IMF distribution is incomplete at the low-mass end such that scaling to the actual IMF may be optimistic and result in underestimating the core efficiency factor.

One observational measure of the BDs in a region is given by the ratio of low-mas stars to BDs: $R=N(0.08-1.0\msun)/(N(0.02-0.08)$. Measurements of local star-forming regions give a range of $R\simeq 2-5$ (e.g. Andersen et al. 2006). For the driven and decaying simulations,  respectively, we find $R > 7$ and $R=3.0$, although these numbers are clearly sensitive to the low-statistics of our simulations. These ratios are most properly represent lower limits  because we have not included the effects of radiative transfer, which have been shown to suppress fragmentation (Krumholz et al. 2007). 
Overall, the driven high-resolution BD mass fraction more closely agrees with the turbulent fragmentation prediction, whereas the undriven BD mass fraction agrees more closely with competitive accretion results, which predict larger BD numbers and mass fractions. Given that BD formation via disk fragmentation dominates in the undriven case, it is unsurprising that these statistics do not agree well with the turbulent fragmentation model and are quite different from one another.

\subsection {Disk Stability}

\begin{figure}
%\plotone{f15c.eps}
\includegraphics[scale=0.5]{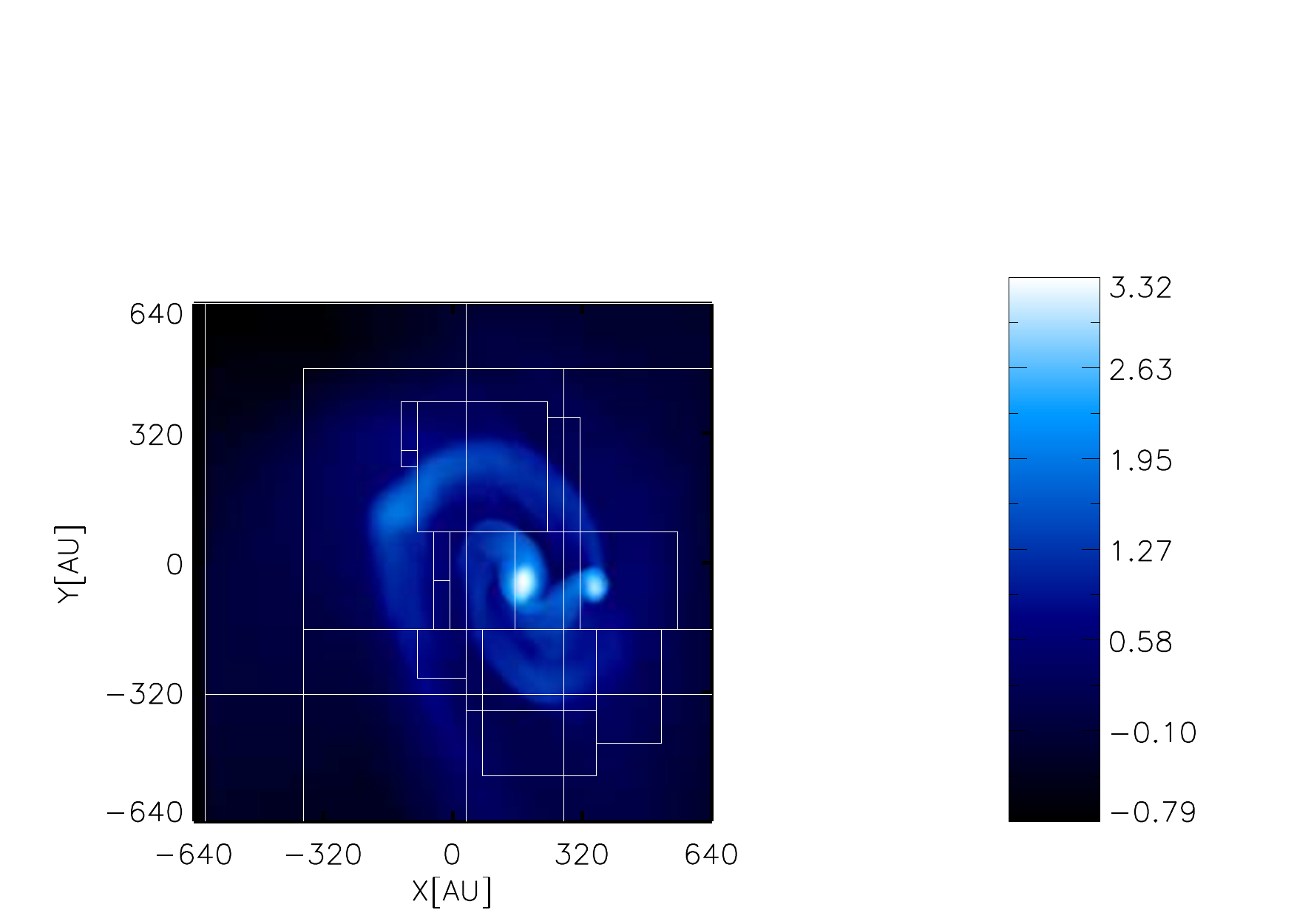}
\figcaption{ The figure shows the log column density (g cm$^{-2}$) of an accretion disk in run D2 with ll levels of refinement. Two fragments have formed at the edges of the spiral disk structure.  The solid black lines denote grid boundaries.
a\label{pltdisk}}		
\end{figure}

Analytically, gravitational disk instability is dictated by the Toomre Q parameter, which is given by
\begin{equation}
Q = {{c_{\rm s} \kappa} \over { \pi G \Sigma}},
\end{equation}
where $\kappa$ is the epicyclic frequency, and $\Sigma$ is the surface density.  For values of $Q \lesssim 1$, the disk becomes unstable to gravitational fragmentation. 
Spiral arms develop for low Q values and fragmentation ensues when Q approaches 1 from above. This fragmentation manifests as a density increase at those locations. The early fragmentation in D2 and U2 generally occurs near the disk perimeters, where it is coldest (e.g. Figure \ref{pltdisk}).  In the simulations, sources of disk instability are due to a combination of perturbations from clumpy infalling gas, gravitational influence of nearby bodies (i.e. other sink particles), and from actual collisions between disks. 
Since the sink accretion radius is $r_{\rm acc}=4 \Delta x$, we neglect the innermost 4 cells in the analysis.  We define the disk gas where $\rho \ge 2 \times 10^{-16}$ g cm$^{-3}$, which agrees fairly well with disk boundaries determined visually. In general, we find disk radii between 150-300AU and surface densities of a few g cm$^{-2}$, values similar to observed properties of low-mass disks (Andrews \& Williams 2006).

The stability of a disk and the onset of gravitational instability have been shown to be correlated with the accretion rate of the disk itself (e.g. Bonnell 1994; Whitworth et al. 1995; Hennebelle et al. 2004; Matzner \& Levin 2005). 
Higher disk accretion rates increase the likelihood of disk instability. This fact agrees with our observation that more instances of disk fragmentation occur in simulations U2 and U3, where there is larger infall onto the clump, in contrast to the case D2 where the disks remain fairly stable.  The level of disk instability is directly visible in the plots of the sink particle accretion rates (Figure \ref{Mdot_plot}); very noisy and irregular accretion corresponds to clumping and disturbance of the disk. The simulations where sinks have many close neighbors show the highest rates of disk instability and episodic accretion.  Note that in Figure \ref{Mdot_plot} of run U4,  the ejection of a companion substantially reduces the accretion rate fluctuations of the remaining protostar.

There has been considerable recent discourse on the necessary criteria for resolving disks and preventing artificial fragmentation (Nelson 2006;  Klein et al. 2007; Vorobyov \& Basu 2005; Durisen et al. 2007).  Since we do find that our disks fragment, this is a topic of concern.
Most recent simulations, including ours, have used the Jeans condition as defined by Truelove et al. (1997) or Bate \& Burkert (1997) to set the minimum refinement of meshes and particles, respectively, in the disk under study. However, Nelson (2006) argues that this criterion is inadequate and inappropriate for cylindrical disk geometry. Additional possible sources of error in our calculation may arise from sink particle gravitational softening, numerical viscosity, and the cartesian nature of the AMR grid. We address these issues here.  

In calculating the gravitational sink particle-particle and sink particle-gas interactions, we use a constant softening length 0.5$\Delta x^{max}$, where $\Delta x^{max}$ is the grid spacing on the maximum level. This is much smaller than both the disk radius and the size of the accretion region, so it should have little effect on the behavior of the disk.  In general, we find that the observed disk fragmentation occurs as cores forming in the ends of spiral arms, well away from the center of the disk (see Figure \ref{pltdisk}). 

Nelson (2006) requires two specific criteria for adequate disk resolution. The first is a Toomre condition,

\begin{equation}
T \geq {{\Delta x^l}\over{ \lambda_{\rm T}}},
\end{equation}
where $T$ is the Toomre number, and $\lambda_{\rm T}$ is the neutral stable wavelength defined by:
\begin{equation}
\lambda_{\rm T} = {{2c_{\rm s}^2}\over{QG\Sigma}}
\end{equation}
and $\Delta x^l$ is the cell spacing on level, $l$.
The above criterion is analogous to the Jeans criterion defined in Truelove et al.,
\begin{equation}
J \geq {{\Delta x^l}\over{ \lambda_{\rm J}}}.
\end{equation} 

\begin{figure}
\plottwo{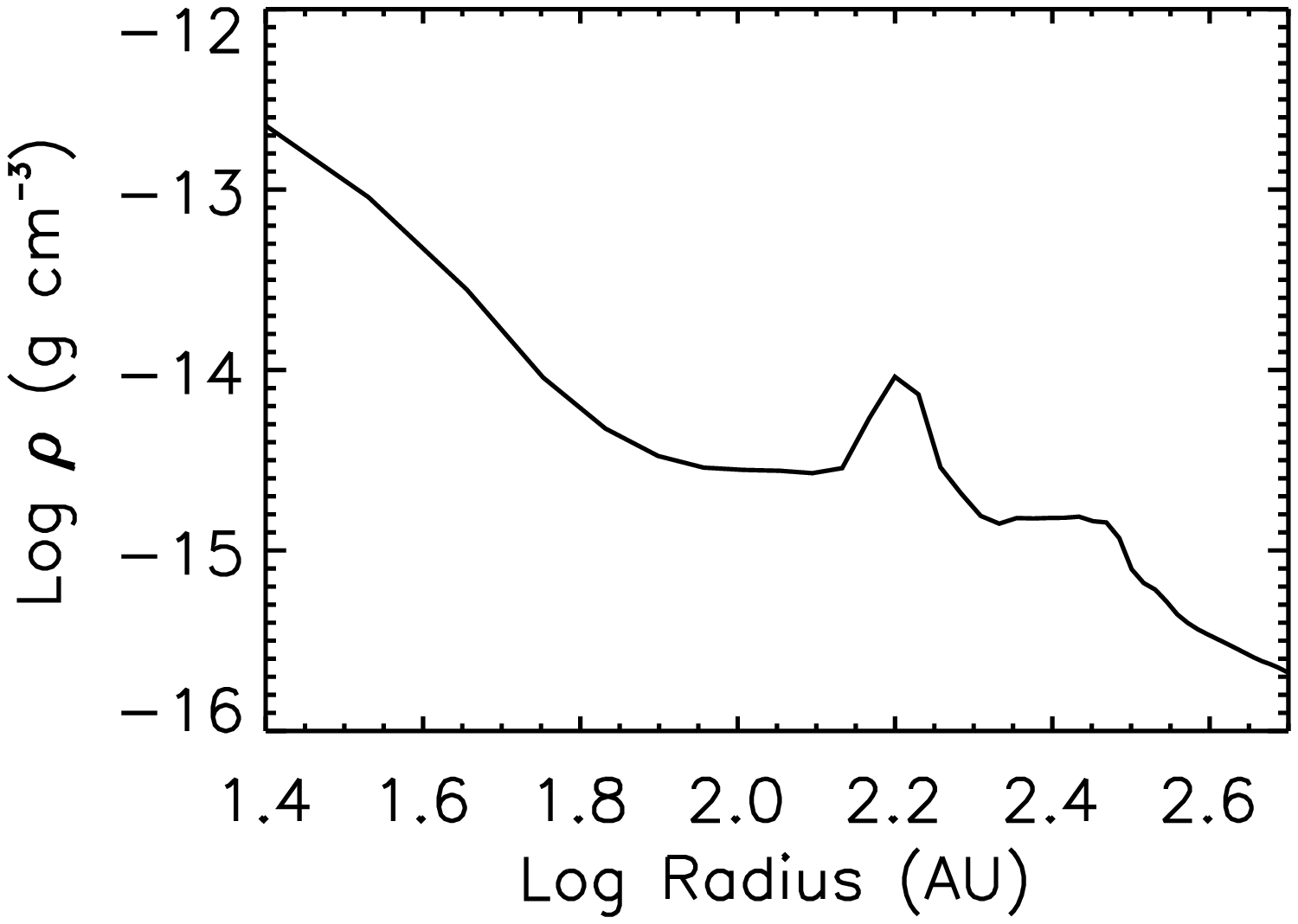}{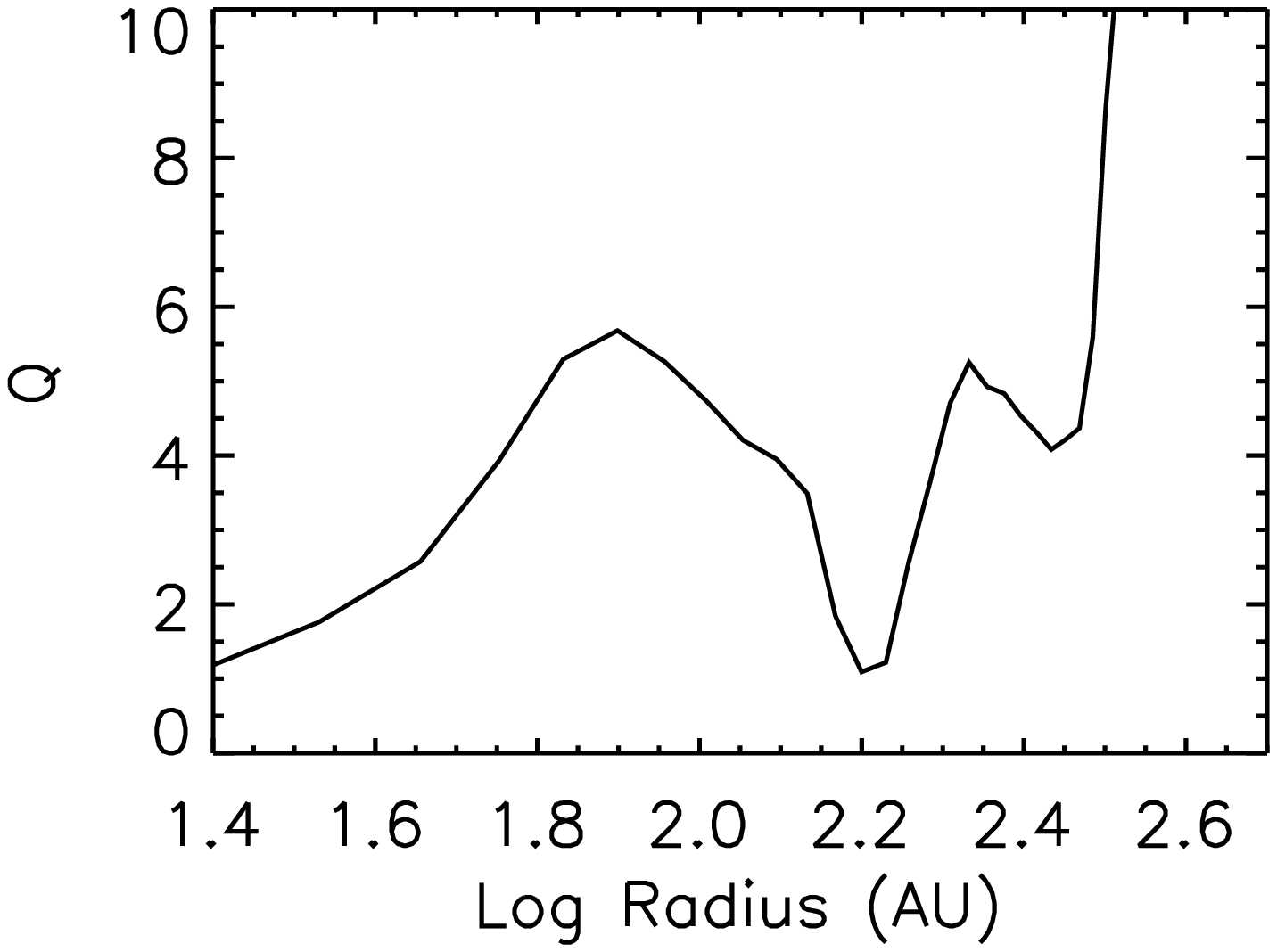}
\plottwo{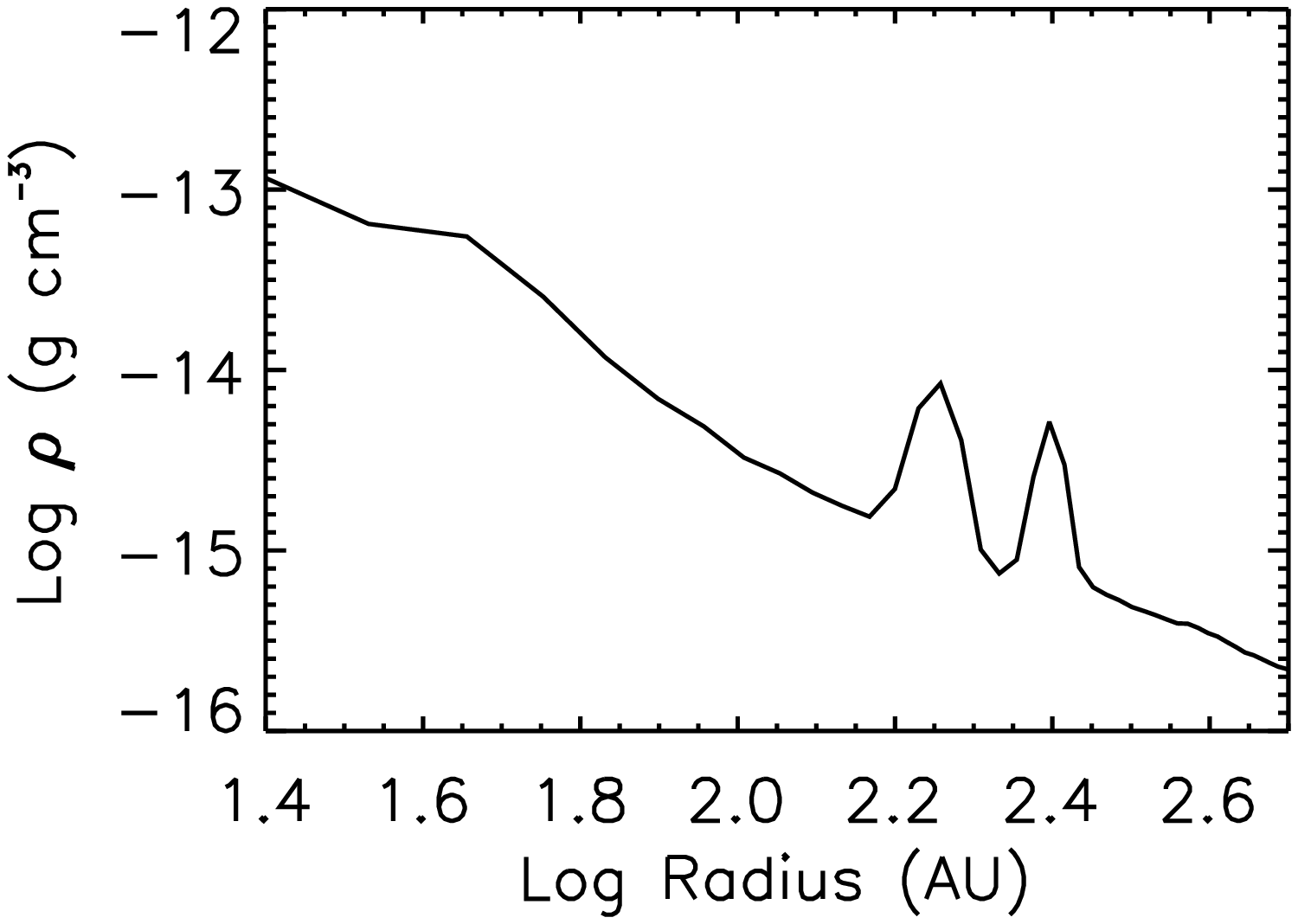}{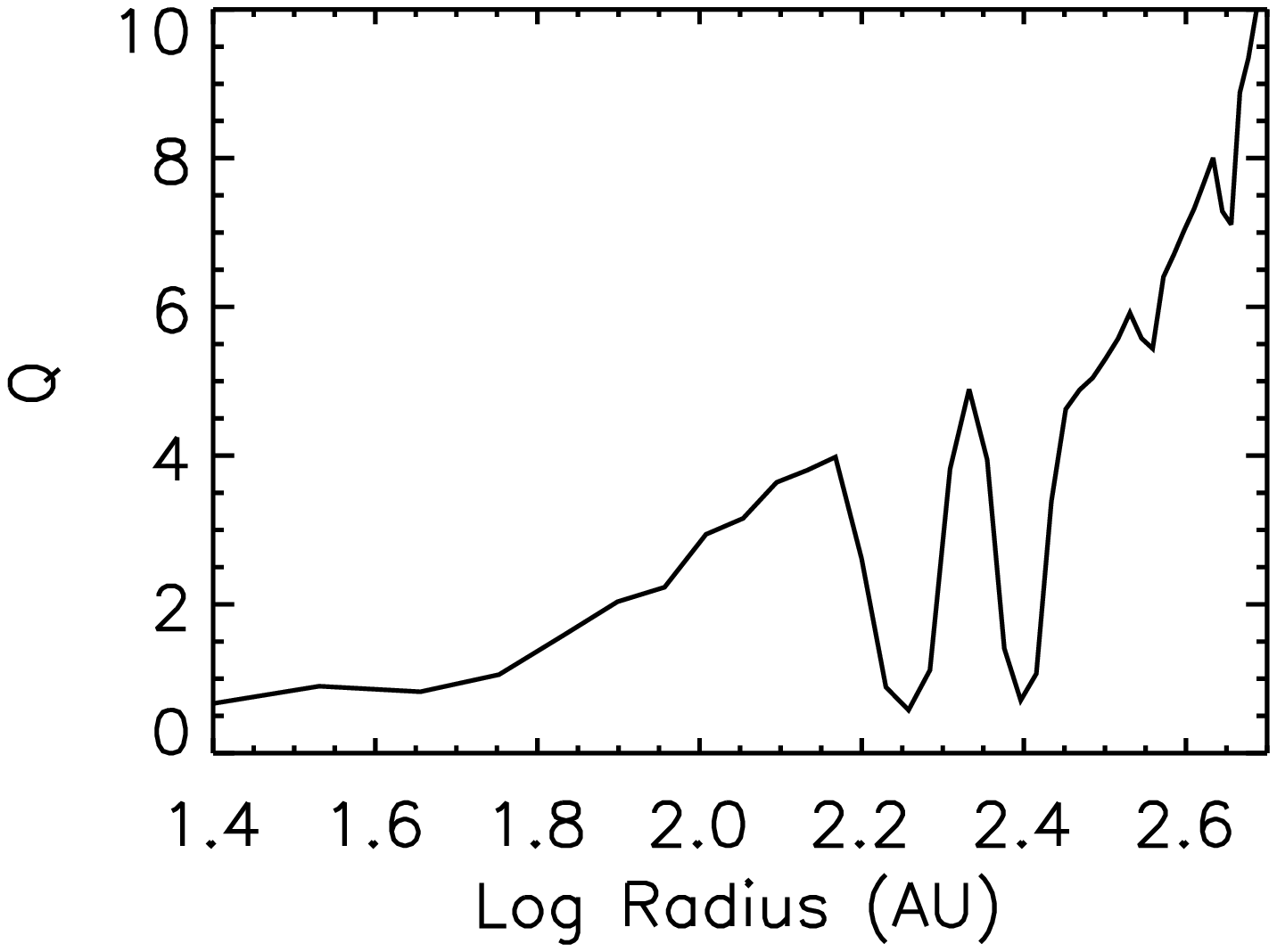}
\plottwo{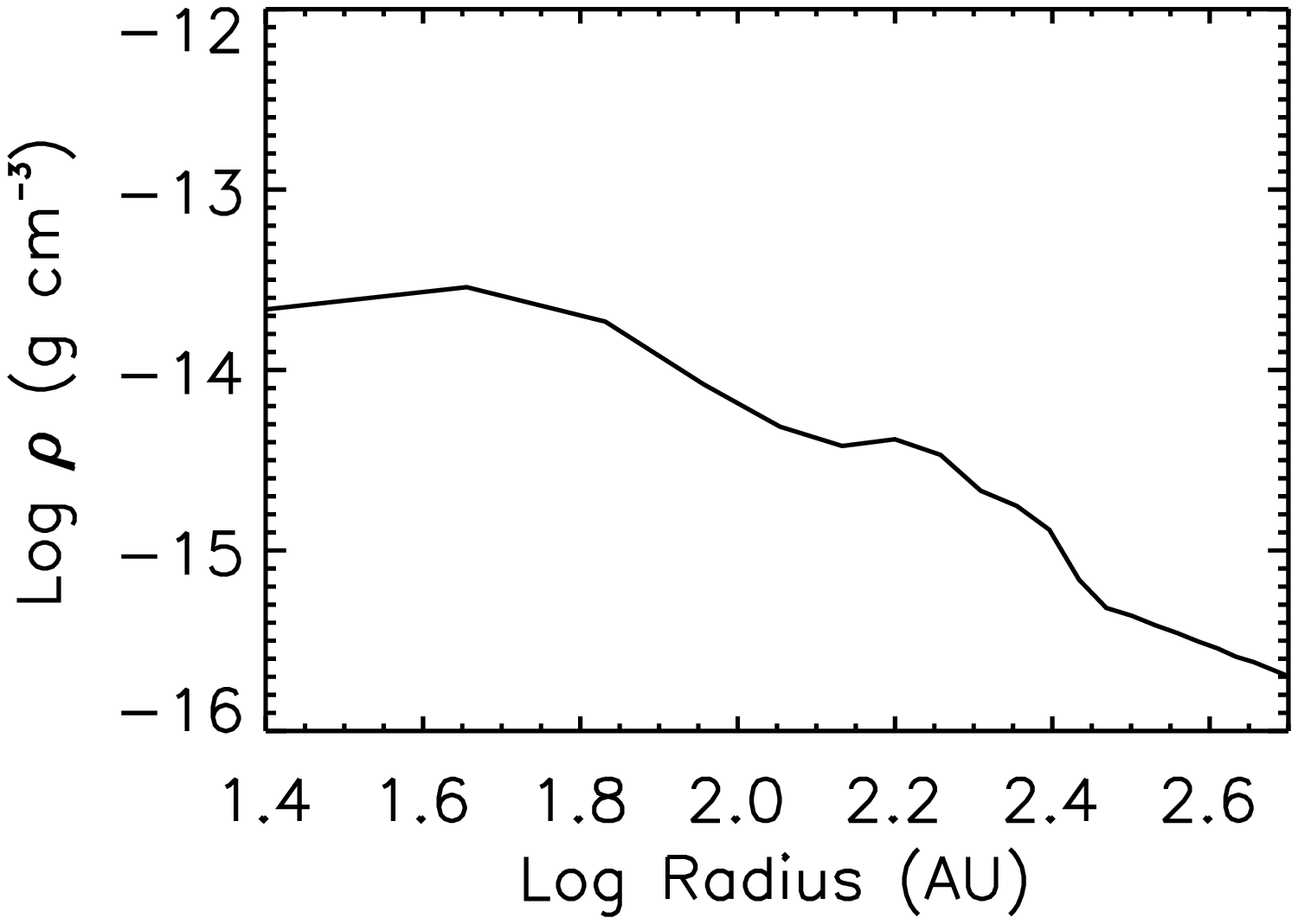}{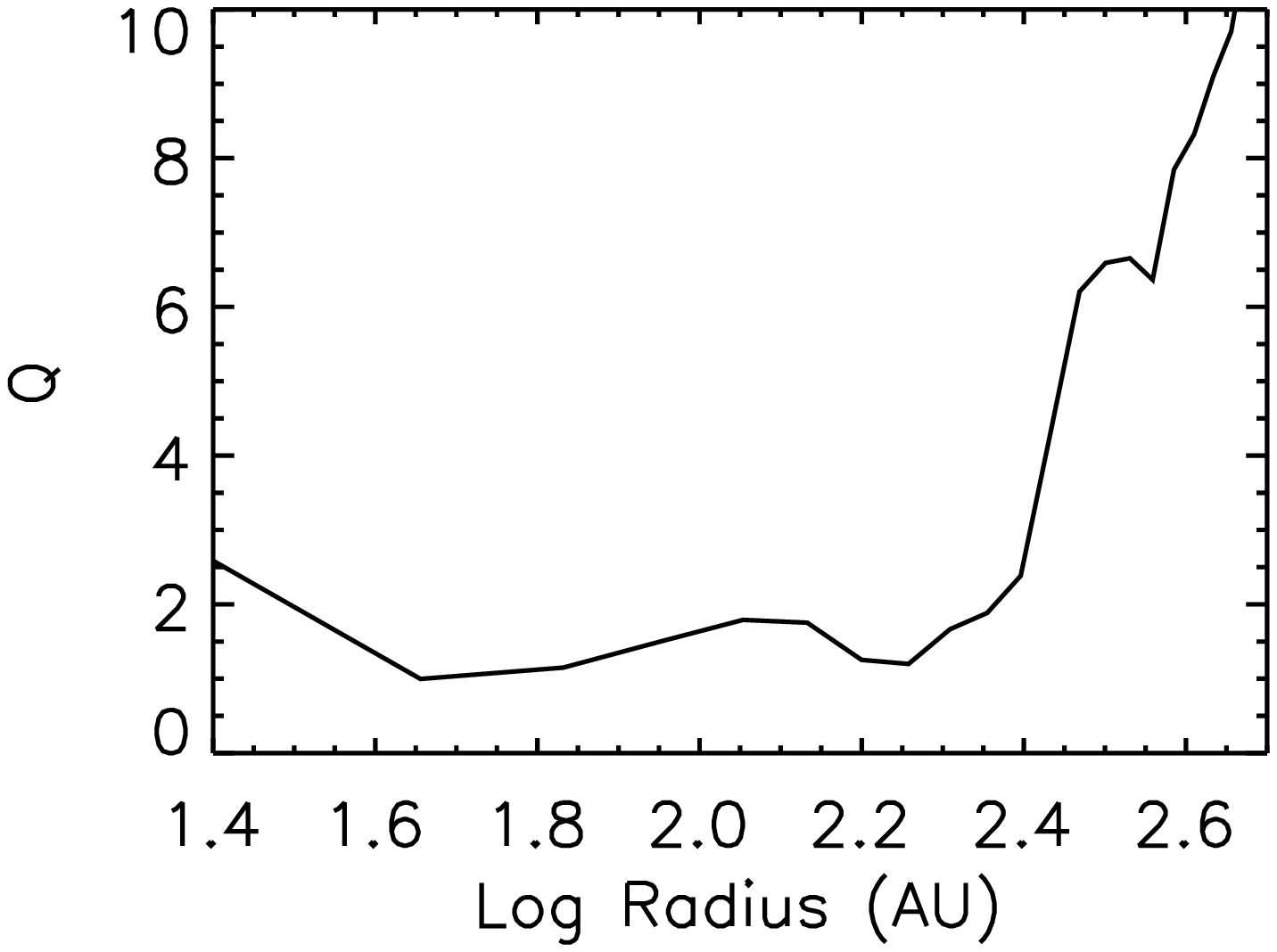}
%{\subfigure{
%\includegraphics[scale=0.5]{f16a.pdf}
%}
%\subfigure{
%\includegraphics[scale=0.5]{f16d.pdf}
%}
%}
%{
%\subfigure{
%\includegraphics[scale=0.5]{f16b.pdf}
%}
%\subfigure{
%\includegraphics[scale=0.5]{f16e.pdf}
%}}
%{
%\subfigure{
%\includegraphics[scale=0.5]{f16c.pdf}
%}
%\subfigure{
%\includegraphics[scale=0.5]{f16f.pdf}}
%}
\figcaption{The figure shows azimuthally averaged disk properties as a function of log radius (AU) for a disk with $\sim$ 2.5 AU (top),  5.0 AU (middle) and 10.0 AU (bottom) resolution. The left plots shows log $\rho$ for the edge on view of the disk. Plots on the right show log $Q$ vs. log $r$.  The central region corresponding to the sink particle accretion region is excluded from the plot. 
\label{Q_res_study}}
\end{figure}

\begin{figure}
\plottwo{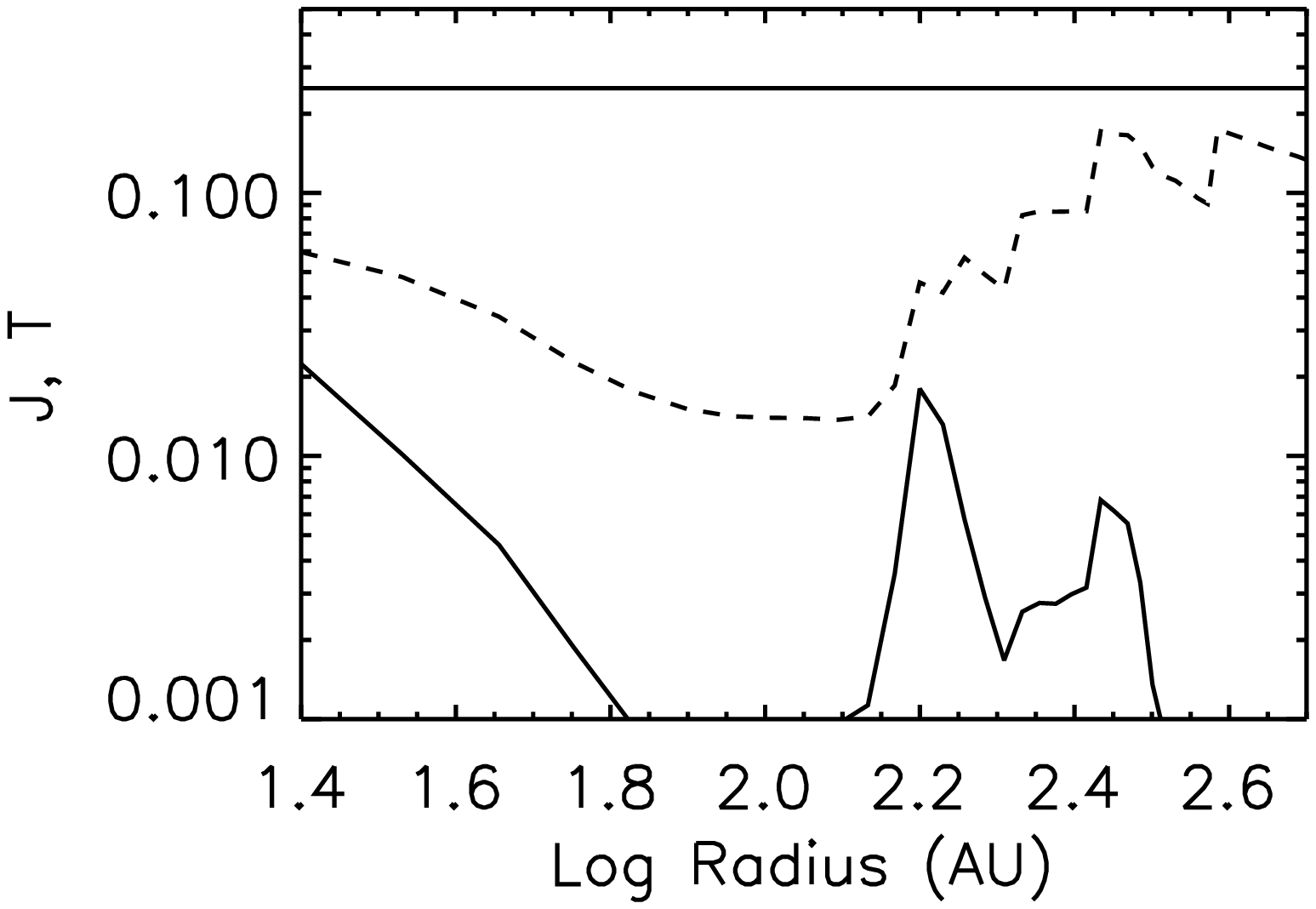}{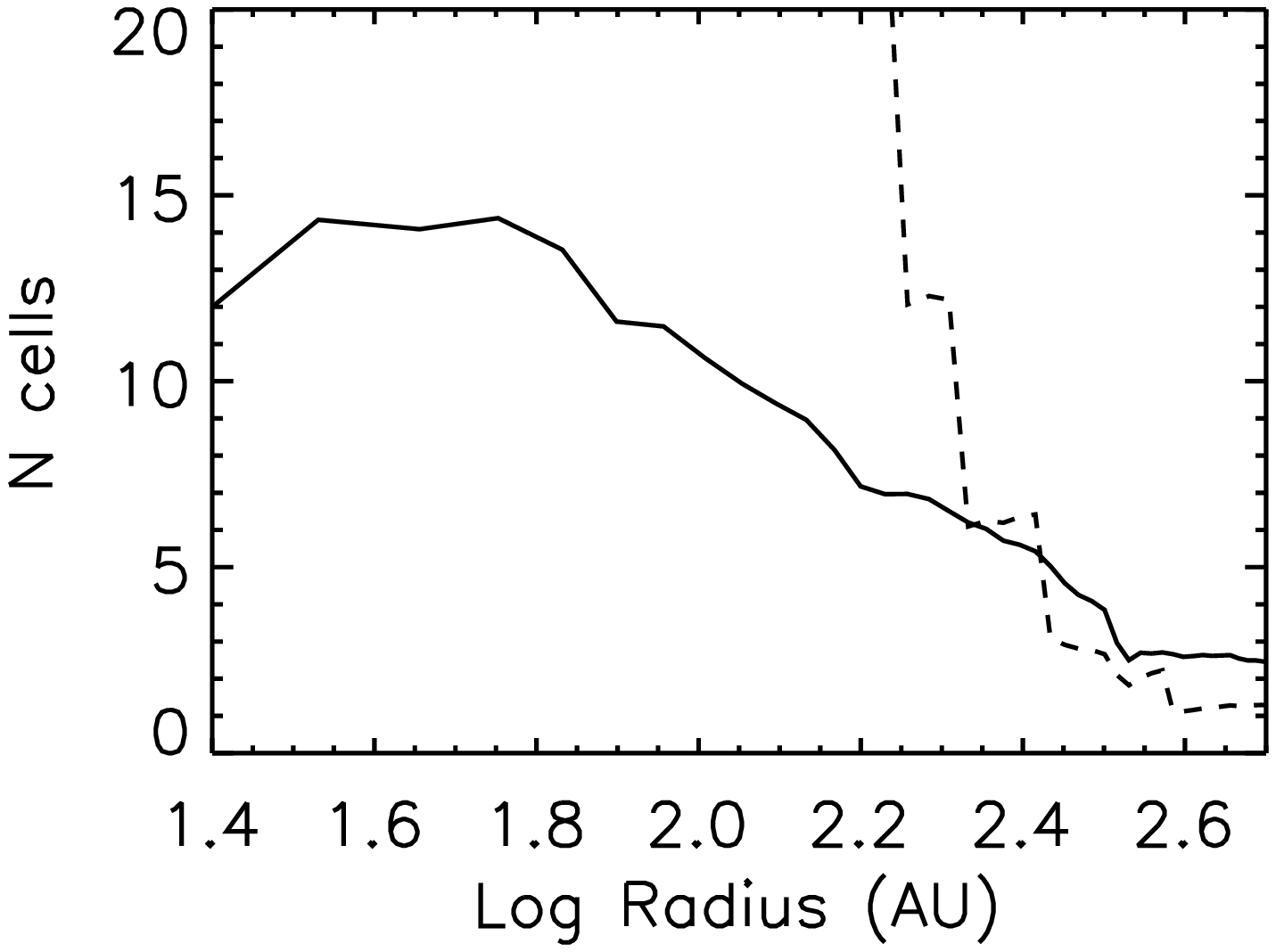}
\plottwo{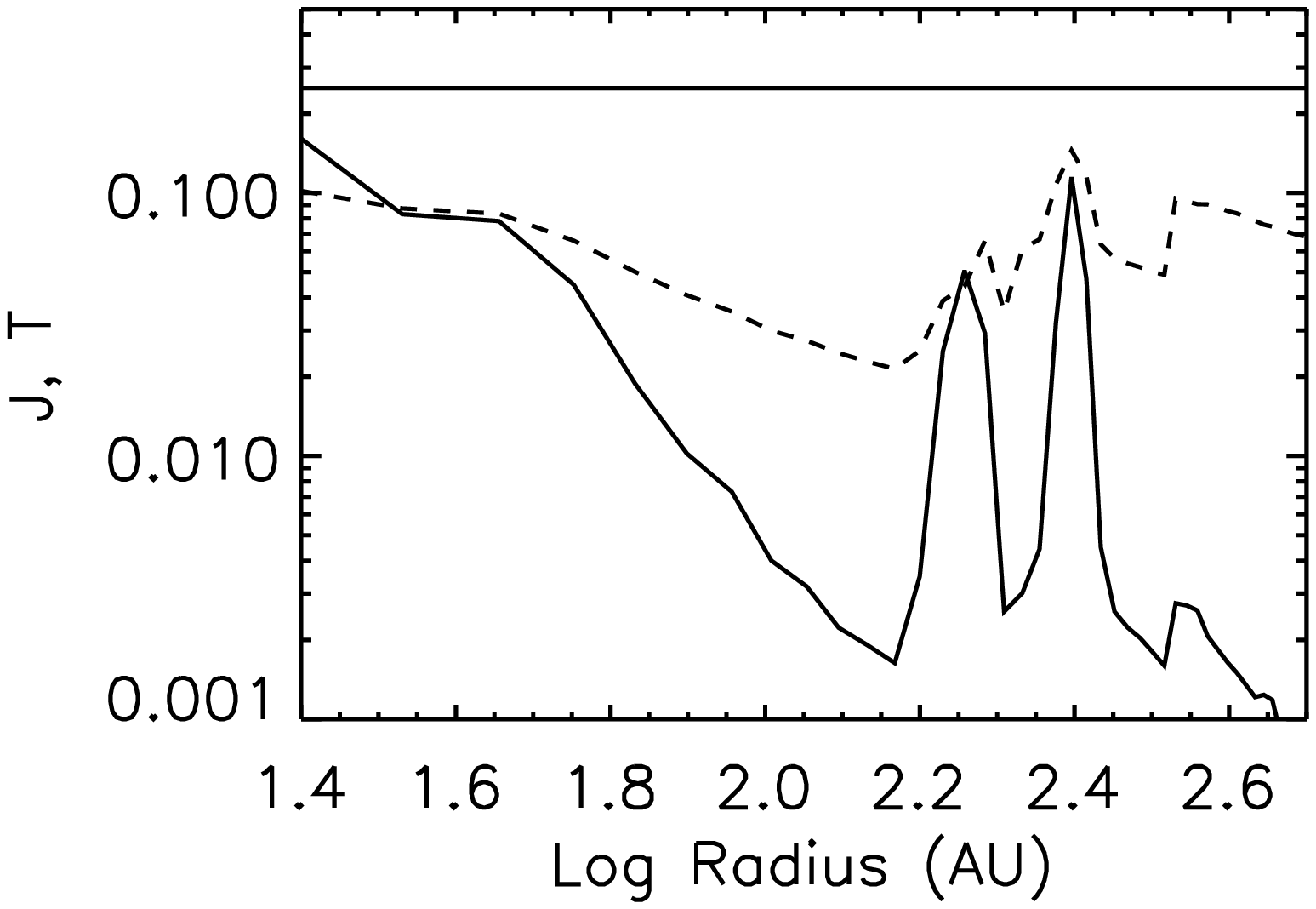}{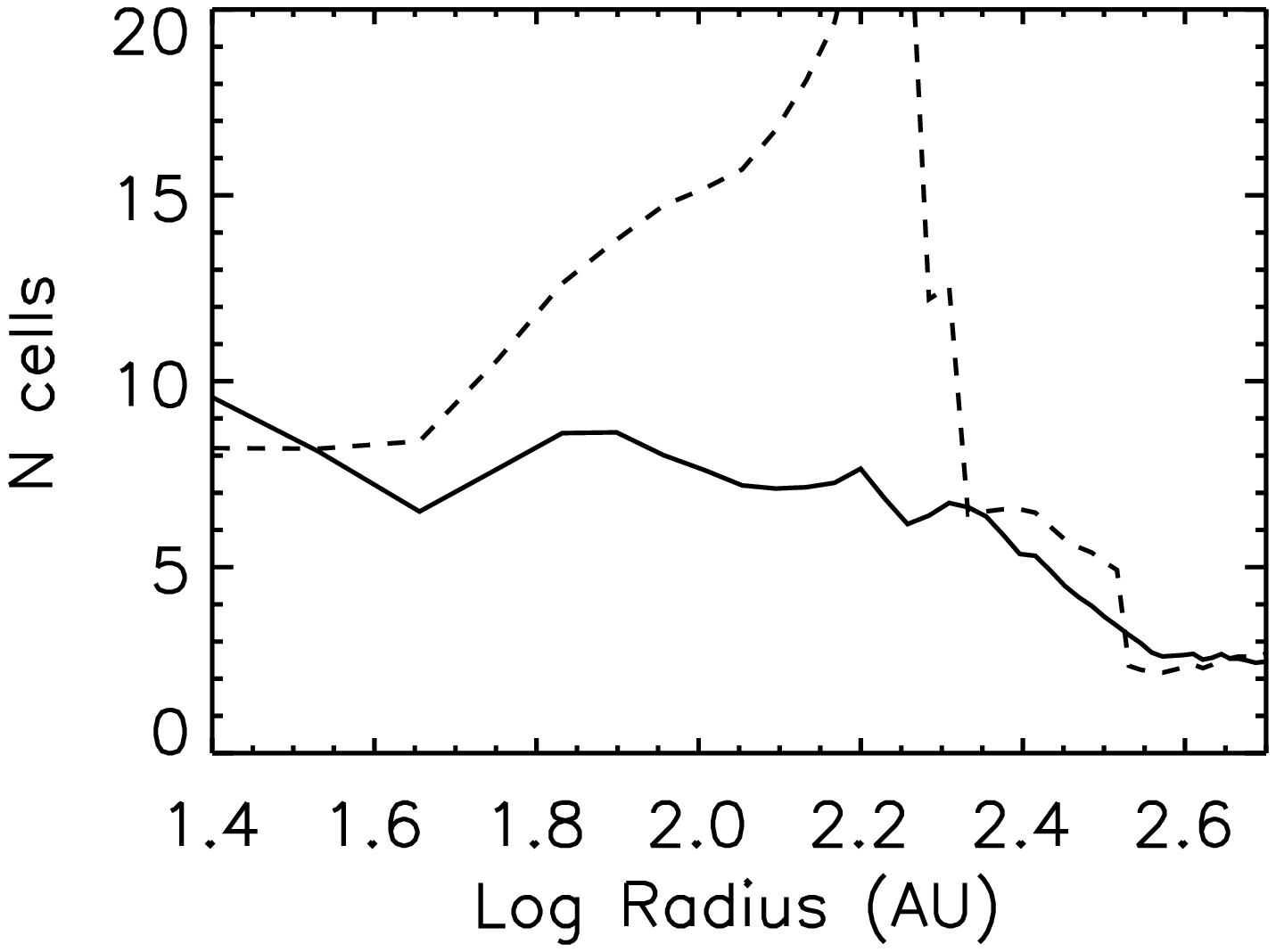}
\plottwo{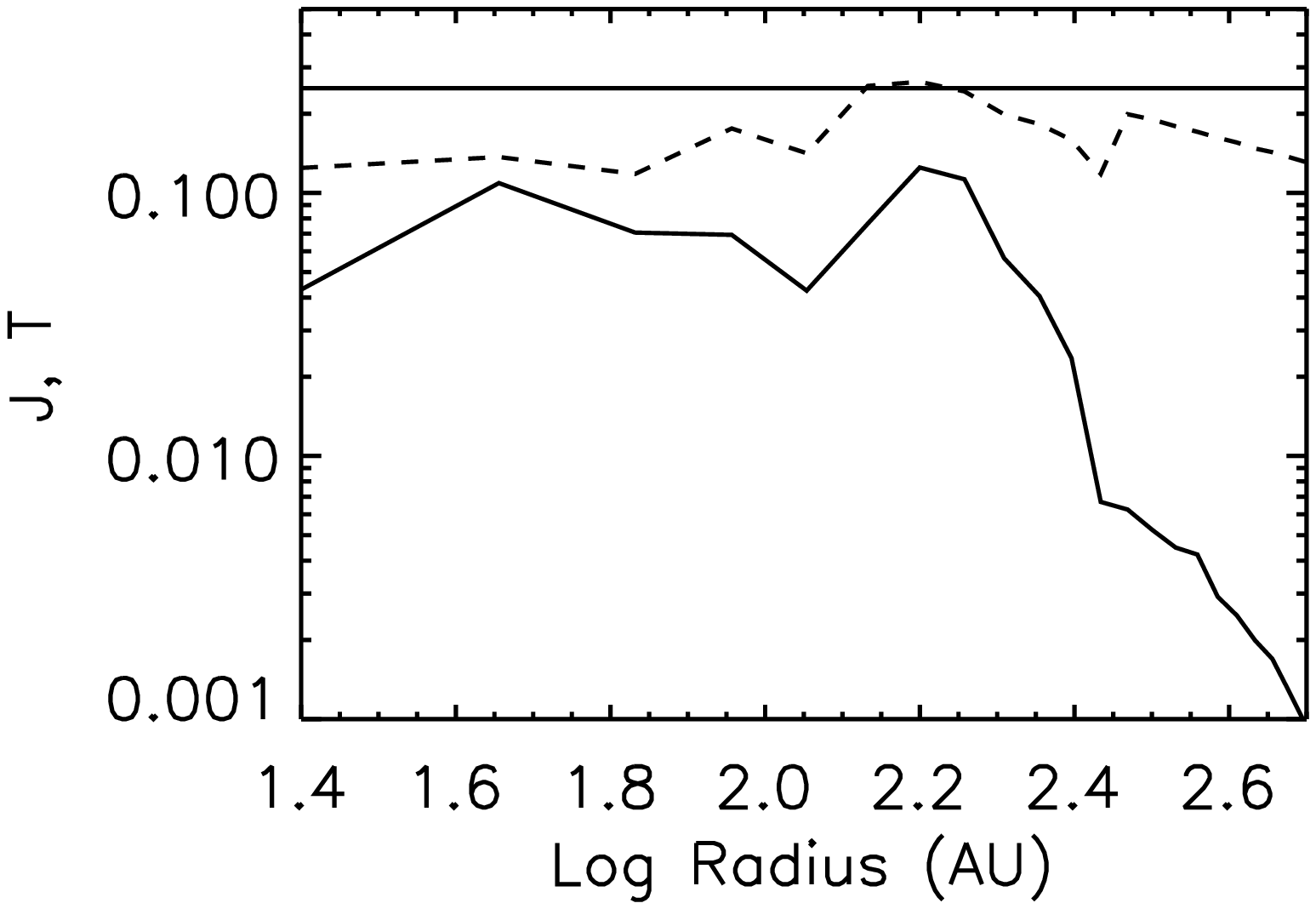}{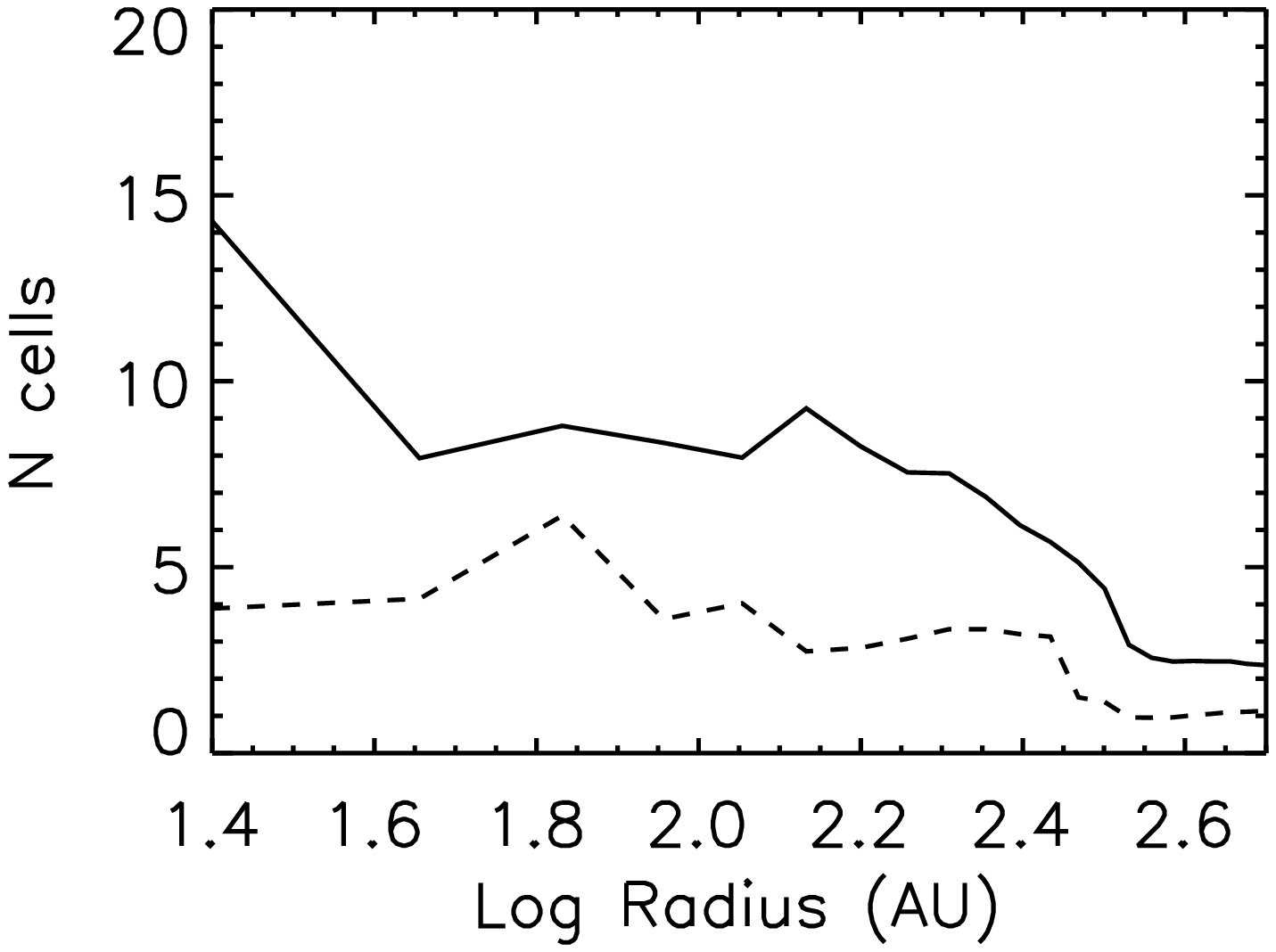}
\figcaption{The figure shows azimuthally averaged disk properties as a function of log radius (AU) for a disk with $\sim$ 2.5 AU (top),  5.0 AU (middle) and 10.0 AU (bottom) resolution. The first column shows plots of  $J$ (dashed line) and $T$(solid line) vs. log $r$, where the horizontal line marks the fiducial value of 0.25. The second column shows the number of cells in the disk vertical scale height as a function of log $r$. The solid line is the required resolution of the vertical scale height according to the Nelson criteria and the dashed line is our resolution.   The central region corresponding to the sink particle accretion region is excluded from the plot.  
\label{disk_res_study}}
\end{figure}

For our simulations, a disk radius of 200 AU is covered by 20 or 40 cells, which is fairly marginal resolution, but we will show it is, in fact, sufficient. We plot the azimuthally averaged density and Toomre Q parameter (equation 13) as a function of radius in Figure \ref{Q_res_study}. Density enhancements are correlated with low Toomre Q in each refinement case. We also plot the right hand sides of equations (14) and (16) as functions of radius in Figure \ref{disk_res_study}.  In all cases, these quantities are under the fiducial value of 1/4. The over resolution in the central part is due to the requirement that all cells surrounding a sink particle be refined to the maximum level in order to encompass the accretion region. 

Figure \ref{pltdisk} indicates the borders between AMR grids, so some disk regions within 200 AU become derefined, and $\Delta x^l \rightarrow \Delta x^{l-1}$.  However, these regions still satisfy the refinement criteria by a good margin.

%SSRO updated 11/5/07, *bw.eps has black and white version

The second criterion formulated by Nelson applies specifically to SPH codes. It postulates the necessity of resolving the disk scale height at the midplane by four smoothing lengths. Nelson argues that insufficient resolution of the vertical structure produces errors in the force balance, thus favoring artificial fragmentation.  If we assume a one-to-one conversion between smoothing lengths and grid cells, we can apply it to our calculation.  Figure \ref{disk_res_study} shows azimuthally averaged quantities for an accretion disk for $\sim$2.5, 5, and 10 AU maximum resolution.  The lowest resolution run fails to adequately resolve the disk scale height, but we do not see extra fragmentation.  We may not see this effect because Nelson formulated and tested his criteria for SPH rather than grid-based codes. It is also possible that the one-to-one analog of smoothing length to $\Delta x$ is not the correct conversion. However, most disagreement is in the inner regions where
the artificial viscosity is high, which suppresses any potential fragmentation. 
In order to determine the cause of the discrepancy, further exploration
with a full grid high resolution study of disks is necessary.

A careful study of the sink particle accretion in a disk is given in Krumholz et al. (2004). For a Keplerian disk, Orion agrees well with analytic results except for some irregularities when the radius of the disk, $ r \sim r_{\rm B}= GM/c_{\rm s}^2 $, the Bondi radius.  However, our simulations have $r << r_{\rm B}$ during the main accretion phase and should be unaffected.  
Also of concern is the magnitude of the numerical viscosity, which has the potential to suppress fragmentation if it is sufficiently high.  Using the definition of $\alpha$ viscosity defined in Krumholz et al. (2004), we can estimate the magnitude of this viscosity as a function of disk radius:
\begin{eqnarray}
\alpha &\simeq& 78 {{r_{\rm B}}\over{\Delta x}} \left( {{r}\over {\Delta x}} \right)^{-3.85} \\
&\simeq&  0.8 M_1 T_{10}^{-1} \Delta x_{5}^{2.85} r_{150}^{-3.85}, 
\end{eqnarray}
where $r_{150}$ is the radial distance from the central star in units of 150 AU, $M_1$ is the stellar mass is units of $M_{\odot}$,  $\Delta x_{5}$ is the cell size in units of 5 AU, normalized to the maximum level of refinement, and $T_{10}$ is the gas temperature in units of 10K.
This expression shows fairly large sensitivity to the cell size and disk radius. Due to the large $\alpha$ value in the inner region of the disk, artificial viscosity is likely to significantly influence the disk properties within the inner 100 AU.  Large values of $\alpha$ could potentially suppress disk fragmentation.
Protostellar disks around low-mass protostars, which are fairly thin and have a low ionization fraction, have been measured to have viscosities of $\alpha \sim 0.01$ (King et. al 2007; Andrews \& Williams 2006).

We find that all disks form exactly two fragments at the same radial locations where $Q\sim 1$.  Convergence of the disk density distribution and number of fragments is our main concern. The averaged quantities are slightly different in the three cases, although the general trends are the same. 
In the lower resolution case the fragmentation is less pronounced,  however two sink particles are introduced at these locations. It is certainly true that the fragments are not well resolved at the lowest resolution, and we are only marginally resolving the disks. Hence we do not devote much discussion in this paper to disk properties. Serious study of disks requires much higher resolution than we adopt in this paper and is best studied
in cylindrical or polar coordinate geometry to minimize the effects of Cartesian cell imprinting.

Given that observations find stellar systems have typically 2-3 stars (Goodwin \& Kroupa 2005), the large number of objects produced in the high resolution decaying simulations seems somewhat anomalous. However, the actual initial multiplicity is more difficult
to determine than multiplicity among field stars due to the difficulty of detecting small
obscured objects, some of which may have separations below the resolvable limit. 
In addition, systems with more than two bodies are unstable and decay through 
gravitational interactions ultimately decreasing the multiplicity of stellar systems over 
time. We witness this behavior in the decaying turbulence protostellar systems, which expel low-mass members.

Nonetheless,  it is likely that a few of these small fragments are numerical products, resulting from our equation of state. For example,  Boss et al. (2000) and Krumholz et al. (2006) both find that fragmentation is sensitive to thermal assumptions and the inclusion of radiative transfer, since heating tends to enhance disk stability. Price \& Bate (2007) show that magnetic fields tend to suppress and delay both fragmentation and spiral disk structure. It is probable that inclusion of radiative feedback and magnetic fields would suppress some of the small objects that we find in the undriven runs. However, the absence of these objects in the driven simulations indicates a striking difference in the accretion rate, system stability, and fragmentation history with turbulent feedback.

%SSRO updated 11/5/07

%SSRO updated 11/5/07

\section {  CONCLUSIONS}

In this paper we use turbulent simulations with AMR to illustrate distinctions between driven and decaying turbulence.  Despite identical initial conditions in the two simulations, we find significant differences between the two cases after one free-fall time.
Our simulations neglect the effect of magnetic fields, which are poorly observationally constrained and occupy a place of ambiguous but potentially large importance (Crutcher 1999). Our simulations also lack radiation transfer, instead relying on the barotropic approximation, which may affect the core fragmentation and protostellar multiplicity in our results. 

We find that the properties of the cores in driven and decaying turbulence at low resolution are not sufficiently different to completely dismiss one turbulent environment. This is in part due to the large scatter in our results. For example, we find that the cores in the different environments have similar shapes and mass-size relations. However, we find that cores in the driven simulation have less rotational energy, which is in better agreement with observations (Goodman et al. 1993; Caselli et al. 2002). The linewidth-size relation of the cores forming in driven turbulence is also closer to the observed relation for low-mass regions (e.g. Jijina \& Adams 1999), while the linewidth-size relation of cores in the decaying simulation is quite flat.  We find that driven turbulence produces a greater number of cores than decaying turbulence with the potential for new star formation occurring longer than a single dynamical time. In contrast, the decaying simulation stops forming new condensations before one global free-fall time. 

The largest differences between the two cases are apparent at high resolution.
We show that our high resolution simulations are converged and that the resolution is sufficient to capture core fragmentation, despite being marginal for determining the detailed properties of disks. We find that the presence or absence of global virial balance has only a subtle influence on individual accretion rate of the largest object forming in the core at least for the first few core free fall times. However, the cores forming in a decaying turbulence environment show clear signs of competitive accretion such that a core's accretion rate is tied to its dynamical history and and its location in the clump. 
This supports the results of Krumholz et al. (2005) who show that simulations exhibiting competitive accretion do so because of lack of a source of turbulence. 

The loss of turbulent feedback in the decaying run affects the dynamic behavior of the forming protostars, resulting in significant disk fragmentation, and BD formation by ejection. This leads to overproduction of BDs in comparison to the observed IMF (e.g. Chabrier 2005).  In contrast, the driven simulations form few BDs, which can be understood in the context of the turbulent fragmentation model for star formation, which predicts BDs to mainly form from small highly compressed cores. Observations of low-mass star forming regions do not find large velocity differences or significant spatial segregation between BDs and low-mass objects as obtained in the decaying simulation.

While our simulations of driven and decaying turbulence
show some statistically significant differences, particularly in the
production of brown dwarfs and core rotation, the uncertainties are large enough
that we are not able to conclude whether observations favor one or the
other. However, in Paper II we use simulated line
profiles to estimate core velocity dispersions and centroid velocities,  and we find that decaying turbulence leads to highly supersonic infall onto protostars, which has not been
observed. Our results thus give some support to the
use of driven turbulence for modeling regions of star formation, but
a conclusive determination of which approach is better awaits larger
simulations with the inclusion of magnetic fields, protostellar outflows
and thermal feedback.
 
 \acknowledgments{
 We thank P.S. Li, M. Krumholz and R. Fisher for helpful discussions and suggestions. 
 Support for this work was provided under the auspices of the US Department
of Energy by Lawrence Livermore National Laboratory under contacts
B-542762 (S.S.R.O.) and DE-AC52-07NA27344 (R.I.K.);  NASA ATP grant NNG06GH96G (CFM and RIK); grant  AST-0606831 (CFM and RIK); and National Science Foundation under
Grant No. PHY05-51164 (CFM and SSRO). Computational resources were provided by the NSF San Diego Supercomputing Center through NPACI program grant UCB267; and the National Energy Research Scientific Computer Center, which is supported by the Office of Science of the U.S. Department of Energy under contract number DE-AC03-76SF00098, though ERCAP grant 80325.}

\clearpage

% Figures created and stored in /volans1/soffner/_clump/_coreanalmbe

% Note the *_0.eps is the black and white version

 %SSRO 10/10/07 updated table

%SSRO 11/6/07 plot_mdot_all2

\end{document}